\documentclass{rmf-d}

\usepackage{nopageno,rmfbib,multicol,times,epsf,amsmath,amssymb,cite}
\usepackage[T1]{fontenc} 
\usepackage[]{caption2}
\usepackage{graphicx}
\usepackage{blindtext}
\usepackage{color}
\usepackage{hyperref}
\usepackage{minted}
\usepackage[numbers]{natbib}
\def\rmfcornisa{Research of Physics \hfill\rmf\ {\bf ??} (*?*) ???--???
\hfill MES? A\~NO?}

\def\rmfcintilla{{\it Rev.\ Mex.\ Fis.\/} {\bf ??} (*?*) (????) ???--???}
\clearpage \rmfcaptionstyle 
\begin{document}
\markboth{Radiation MHD chromospheric jet}{ A \LaTeX template for the RMF}
%
%
\title{Radiation magnetohydrodynamics modeling of an impulsively driven chromospheric jet in the solar atmosphere
\vspace{-6pt}}
\author{J. J. Gonz\'alez-Avil\'es}
\address{Escuela Nacional de Estudios Superiores (ENES) Unidad Morelia, Universidad Nacional Aut\'onoma de M\'exico, Unidad Morelia 58910, Morelia, Michoac\'an, M\'exico}
%
%
\maketitle
%
%
\begin{abstract}
\vspace{1em} 
%
%
In this paper, we present a numerical simulation of an impulsively driven chromospheric jet in the solar atmosphere using the non-ideal magnetohydrodynamic (MHD) equations coupled with frequency- and angle-averaged radiation transport equations. These include the dynamics of the radiation energy density and radiation flux. The jet is initiated by a localized Gaussian pulse applied to the vertical velocity component in the upper chromosphere ($y = 1.75$ Mm), producing a collimated plasma structure that exhibits characteristics similar to macrospicules. We focus on the formation and evolution of the chromospheric jet as it propagates through an optically thin region encompassing the upper chromosphere and solar corona, where both the Planck-averaged absorption and Rosseland-averaged scattering opacities are low. Although radiation transport terms only slightly affect the jet’s morphology, they play a significant role in governing radiative processes in the corona. In particular, radiation transport contributes to the dissipation of the chromospheric jet, which effectively acts as a radiative cooling mechanism as the jet evolves through the optically thin solar corona. 
\vspace{1em}
\end{abstract}
\keys{ \bf{\textit{
Solar atmosphere -- Solar corona -- Magnetohydrodynamics -- Radiative magnetohydrodynamics -- Computational methods
}} \vspace{-8pt}}
\begin{multicols}{2}

\section{Introduction}
\label{sec:Introduction}

Jets are dynamic plasma ejections in the solar atmosphere, widely observed in both quiet and active regions across multiple wavelengths, including X-rays, Extreme Ultraviolet (EUV), and H$\alpha$. To explain these observations, several physical mechanisms have been proposed, both theoretically and numerically, with magnetic reconnection being one of the most widely accepted jet triggering processes \cite{Shibataetal2007, Gonzalez-Avilesetal2017, Gonzalez-Avilesetal2018}. Among the various types of jets, spicules are particularly notable. They are thin spike-shaped plasma structures that are predominantly observed in the chromosphere \cite{Beckers1972}. Depending on their size and lifetime, spicules are commonly classified into three types: (i) Type I spicules, with lengths of 7–11 Mm, lifetimes of 5–15 minutes, and upward velocities around 25 km s$^{-1}$ \cite{1968SoPh....3..367B,2009SSRv..149..355Z};
(ii) Type II spicules, with average heights of ~5 Mm, propagation speeds of 50–100 km s$^{-1}$, and short lifetimes ranging from 10 to 150 seconds \cite{10.1093/pasj/59.sp3.S655, Sekse_et_al_2012};
(iii) Macrospicules, typically observed in polar coronal holes, reach heights between 7 and 70 Mm above the solar limb, with maximum velocities from 10 to 150 km s$^{-1}$ and lifetimes ranging from 3 to 45 minutes \cite{1975ApJ...197L.133B, Loboda_2019}.

In the context of numerical MHD simulations, a chromospheric jet that exhibits spicule-like features, such as height, vertical velocity, and lifetime, can be modeled as an impulsively driven jet. This type of jet is a dynamic plasma outflow initiated by a spatially localized perturbation (e.g., a pressure pulse, reconnection burst, or velocity pulse) in a magnetized plasma. The resulting disturbance generates a high-speed, narrow jet of material that propagates along magnetic field lines in the solar atmosphere. Several MHD simulations of impulsively driven chromospheric jets reproduce spicule-like characteristics, including 2D adiabatic, non-adiabatic, and two-fluid models \citep[see, e.g.,][]{Murawski&Zaqarashvili2010, Murawskietal2011, Kuzmaetal2017, Singhetal2019, Gonzalez-Avilesetal2021, Gonzalez-Aviles_et_al_2022, Srivastavaetal2023}. In all these studies, the jet formation mechanism is based on the classical rebound shock model \citep{Sterling&Hollweg1988, Sterling&Mariska1990}, which explains spicule formation as the result of shock waves generated by localized impulsive disturbances. These shocks propel chromospheric plasma upward along magnetic field lines, producing short-lived, narrow jets that closely resemble Type I spicules in both morphology and dynamics. In contrast, the formation of Type II spicules is not attributed to impulsive driving. Instead, 2D and 3D MHD simulations indicate that their generation involves more complex physical processes, including magnetic resistivity, thermal conduction, radiative transfer, partial ionization, and ambipolar diffusion \citep{Gonzalez-Avilesetal2017, Gonzalez-Avilesetal2018, Martinez-Sykoraetal2017a, Martinez-Sykoraetal2017b, DePontieuetal2017}. Moreover, the numerical modeling of macrospicules convergence on two primary formation mechanisms: pulse- or shock-driven models \citep{Muraswkietal2011, Kayshap_et_al_2013, Gonzalez-Avilesetal2021} and reconnection-driven models \citep{Kamio_et_al_2010, Raouafi_et_al_2016, Duan_et_al_2023}. Complementary observational inversions in He II 304 \text{\AA} provide critical constraints, revealing non-ballistic shock-driven dynamics with typical velocities of 70–140 km s$^{-1}$ and pronounced decelerations \citep{Loboda_2019}.

Despite substantial progress in the modeling of impulsively driven chromospheric jets, such jets have not yet been simulated using magnetohydrodynamic (MHD) equations coupled with a system of frequency and angle averaged radiation transport equations that independently evolve both the radiation energy density and the three components of the radiation flux. In this work, we present a numerical simulation of an impulsively driven jet in the solar atmosphere based on non-ideal MHD equations that incorporate highly anisotropic thermal conduction and empirical coronal heating. These equations are coupled to a radiation transport system that evolves both the radiation energy density and the flux. The simulation focuses on the formation and evolution of a jet propagating through a region that includes the upper chromosphere and the solar corona, a medium characterized by low Planck-averaged absorption opacity and low Rosseland-averaged scattering opacity. In the low Planck mean opacity regime, radiative coupling via emission and absorption is weak, while in the low Rosseland mean opacity regime, photon diffusion is efficient and radiative transport is enhanced. In addition, we investigate whether the inclusion of radiation transport affects the morphology and dynamics of the chromospheric jet. This is motivated by previous models \citep{Sterling&Mariska1990, 2000SoPh..196...79S, Gonzalez-Avilesetal2021} that incorporated optically thin radiative cooling functions into the MHD equations and found that such effects can influence the dynamics of chromospheric jets with properties similar to those of spicules. 

The paper is organized as follows. In Section \ref{sec:Model_num_methods}, we describe the system of radiation MHD equations, including the relevant source and radiation terms. We also outline the solar atmospheric model, magnetic field configuration, radiation parameters, and the applied perturbation. In addition, we describe the details of the numerical methods used to solve the radiation MHD equations. In Section \ref{sec:results_num_simulations}, we present the results of the numerical simulations. Section \ref{sec:discussion} discusses the impact of the radiation transport terms on jet dynamics, as well as the influence of source terms on the background solar atmosphere. Finally, Section \ref{sec:conclusions_and_final_comments} summarizes the conclusions and provides the final remarks.

\section{Model and Numerical Methods}
\label{sec:Model_num_methods}

\subsection{The system of radiation MHD equations}
\label{sub-sec:System_equations}

We consider a gravitationally stratified solar atmosphere described by a plasma coupled with the radiation field, for which a two-moment approach is adopted under the gray approximation. The resulting radiation magnetohydrodynamics (MHD) equations can be written in quasi-conservative form as   
\begin{align}
\frac{\partial\rho}{\partial t} + \nabla\cdot(\rho{\bf v}) = 0 \label{mass_cons_eq} \\
\frac{\partial{(\rho\bf v})}{\partial t} + \nabla\cdot\left[{\rho\bf v\otimes\bf v}-{\bf B\otimes \bf B} + {\bf I}\left(p+\frac{{\bf B}^{2}}{2}\right)\right]^{T} = \rho{\bf g} + {\bf G} \label{momentum_cons_eq} \\
\frac{\partial E_{t}}{\partial t} + \nabla\cdot\left[\left(\frac{\rho{\bf v}^{2}}{2}+\rho e + p\right){\bf v} + c{\bf E}\times{\bf B}\right] = {\rho\bf v\cdot g} + cG^{0}  \nonumber \\ +\nabla\cdot{\bf F}_{c} + H,  \label{energy_cons_eq} \\
\frac{\partial{\bf B}}{\partial t} + \nabla\times{c{\bf E}} = 0 \label{induction_eq} \\
\frac{1}{\hat{c}}\frac{\partial E_{r}}{\partial t} + \nabla\cdot{\bf F}_{r} = -G^{0} \label{energy_rad_eq} \\
\frac{1}{\hat{c}}\frac{\partial{\bf F}_{r}}{\partial t} + \nabla\cdot\mathbb{P}_{r} = -{\bf G}, \label{flux_rad_eq}
\end{align}
where $\rho$, $p$, ${\bf v}$, $e$, ${\bf B}$, and ${\bf E}$ are the plasma density, pressure, internal energy, velocity field, magnetic and electric fields. While $E_{r}$, ${\bf F}_{r}$, and $\mathbb{P}_{r}$ are the radiation energy, radiation flux, and radiation pressure tensor, respectively. There, ${\bf I}$ represents the identity matrix, $c$ is the speed of light, and $\hat{c}$ is the reduced speed of light, which is introduced to reduce the overhead of the explicit integration of radiation transport terms. In sources terms, the equations include the gravitational acceleration vector ${\bf g} = [0, -g]$ with magnitude $g=2.74\times10^{4}$ cm s$^{-2}$, the thermal conduction flux ${\bf F}_{c}$, the coronal heating function $H$, and the radiation interaction terms defined by $(G^{0}, {\bf G})$. In addition, $E_{t}$ denotes the total energy density, i.e., the sum of the internal, kinetic, and magnetic energy densities,  

\begin{center}
\begin{equation}
    E_{t} = \rho e + \frac{\rho{\bf v}^{2}}{2} + \frac{{\bf B}^{2}}{2},
\end{equation}
\end{center}
where a thermally ideal gas provides the closure $\rho e = \rho e(p,\rho)$, in terms of the following equation of state (EOS)

\begin{equation}
    p = \frac{\rho k_{B} T}{m_{u}\mu},
\end{equation}
where $k_{B}$ is the Boltzmann constant, $m_{u}$ is the atomic mass unit and $\mu$ is the mean molecular weight. This paper uses a value of $\mu = 0.6$, representing a fully ionized plasma, that is, a gas assumed about 92\% of H and 8\% of He.

The contribution from thermal conduction ${\bf F}_{c}$ is defined in terms of a flux-limited that smoothly varies between the classical and saturated thermal conduction regimes $F_{\textrm{class}}$ and $F_{\textrm{sat}}$, respectively, that is: 

\begin{equation}
   {\bf F}_{c} = \frac{F_{\textrm{sat}}}{F_{\textrm{sat}}+|{\bf F}_{\textrm{class}|}}{\bf F}_{\textrm{class}}. \label{thermal_conduc_flux}
\end{equation}
In the above equation, the thermal conduction flux is highly anisotropic along the direction of the magnetic field lines, that is, 

\begin{equation}
    {\bf F}_{\textrm{class}} = \kappa_{\parallel}\frac{{\bf B}({\bf B}\cdot\nabla T)}{|{\bf B}|},  
\end{equation}
where $\kappa_{\parallel} = 8\times10^{-7}$ erg s$^{-1}$ cm$^{-1}$ K$^{-1}$, which represents a typical value in the solar corona \cite{Ruan2020}. In the saturated flux limit, $F_{\textrm{sat}} = 5\phi\rho c_{\textrm{iso}}^{3}$, where $c_{\textrm{iso}}$ is the isothermal speed of the sound and $\phi = 0.3$ is a free parameter that helps fix the value of $F_{\textrm{sat}}$ \cite{1984ApJ...277..605G}.

The coronal heating function $H$ that contributes as a source term to the total energy equation (\ref{energy_cons_eq}) helps to maintain a background solar corona against thermal conduction and radiative losses, as given in \cite{Ruan2020} 

\begin{equation}
H = \textrm{max}\left(\frac{c_{0}*[(y-y_{0})/h_{0}]^{-2/7}}{\exp[h_{1}/(y-y_{1})]-1},0\right),
\end{equation}
where $c_{0}=0.01$ erg cm$^{-3}$ s$^{-1}$, $y_{0} = 1$ Mm, $h_{0} = 5$ Mm, $y_{1} = 2$ Mm and $h_{1}=3$ Mm. This background empirical heating function reaches its maximum value in about the transition region $\sim 2.1$ Mm, where it compensates for the cooling that radiation source terms could produce.

The components of the radiation pressure tensor, $\mathbb{P}_{r}$, are defined as a function of $(E_{r}, {\bf F}_{r})$ through the closure of M1 introduced by \cite{LEVERMORE1984149}, as 

\begin{equation}
P^{ij} = E_{r}\left(\frac{1-\xi}{2}\delta^{ij}+\frac{3\xi-1}{2}n^{i}n^{j}\right), \quad \xi = \frac{3+4f^{2}}{5+2\sqrt{4-3f^{2}}}, 
\end{equation}
where ${\bf n} = {\bf F}_{r}/|{\bf F}_{r}|$, $f=|{\bf F}_{r}|/E_{r}$, the sub-indexes $i,j$ denote the spatial components of the radiation pressure tensor $P^{ij}$, and $\delta^{ij}$ is the Kronecker delta. The radiation-matter interaction terms $(G^{0}, {\bf G})$ are defined by boosting into the Eulerian frame their co-moving values, given by 
\begin{equation}
(\tilde{G}^{0}, \tilde{{\bf G}})_{\mathrm{comov}} = \rho[\kappa(\tilde{E_{r}}-a_{R}T^{4}), (\kappa + \sigma)\tilde{{\bf F}}_{r}]_{\mathrm{comov}}. 
\end{equation}

In this equation, $\kappa$ and $\sigma$ are the frequency-averaged absorption and scattering coefficients, respectively, $a_{R} = 4\sigma_{\mathrm{SB}}/c$ is the radiation constant, and $T$ is the fluid temperature. The interaction terms in the laboratory frame that are the actual source terms in the equations are obtained by the following Lorentz transformation laws to the first order in ${\bf \beta}= {\bf v}/c$: 

\begin{eqnarray}
    G^{0} = \tilde{G}^{0} + {\bf \beta}\tilde{{\bf G}} \nonumber \\
    {\bf G} = \tilde{{\bf G}} + \tilde{G}^{0}{\bf \beta}. 
\end{eqnarray}

Similarly, the radiation fields are transformed into the laboratory frame to the first order in $\beta$, as 

\begin{eqnarray}
    E_{r} &=& \tilde{E}_{r} + 2\beta_{i}\tilde{F}_{i} \nonumber \\
    F_{r}^{i} &=& \tilde{F}_{r}^{i} + \beta^{i}\tilde{E}_{r} + \beta_{j}\tilde{P}_{r}^{ij} \nonumber \\
    P_{r}^{ij} &=& \tilde{P}_{r}^{ij} + \beta^{i}\tilde{F}^{j} + \beta^{j}\tilde{F}^{i}. 
\end{eqnarray}

This implies the following expressions for the source terms implemented in the code: 

\begin{eqnarray}
    G^{0} = \rho\kappa(E_{r}-a_{R}T^{4}-2\beta\cdot{\bf F}_{r}) \nonumber \\ + \rho\chi\beta\cdot({\bf F}_{r}-E_{r}\beta-\beta\cdot\mathbb{P}_{r}) \nonumber \\
    {\bf G} = \rho\kappa(E_{r}-a_{R}T^{4}-2\beta\cdot{\bf F}_{r})\beta \nonumber \\
    + \rho\chi({\bf F}_{r}-E_{r}\beta-\beta\cdot\mathbb{P}_{r}),
\end{eqnarray}
where some $\mathcal{O}(\beta^{2})$ terms are kept in order to recover the local thermal equilibrium (LTE) limit given by $\tilde{E}_{r}\to a_{R}T^{4}$ and $\tilde{F}_{r}\to{\bf 0}$ when $\sigma,\kappa\to0$ \cite{Jiangetal2012}.  

We can summarize the pros and cons of the closures employed in the radiation subsystem: i) The Eddington approximation assumes an isotropic radiation that works for optically thick regimes and provides a smooth transition from diffusion-like to free-streaming behavior. Nevertheless, it can be inaccurate in optically thin regions and poor in highly anisotropic cases. ii) The flux-limited diffusion (FLD) closure introduces a flux limiter to handle transitions between optically thick and thin regimes, so it ensures a smooth transition from diffusion (optically thick) to free-streaming (optically thin) behavior and avoids unphysical superluminal fluxes. Nevertheless, it can be inaccurate in anisotropic fields and might lead to excessive diffusion. iii) The exact Eddington tensor solves the full radiative transfer equation (non-angular-averaged) to obtain the Eddington tensor, which makes it the most accurate of the three closures since it accounts for angular anisotropy. Also, it captures both optically thick and thin regimes without empirical flux limiters. Still, it is computationally expensive because it requires solving the full angular-dependent radiative transfer equation and becomes complex for multidimensional problems.

\subsubsection{The model of the solar atmosphere}
\label{sub-sec:Model_solar_atmosphere}

\begin{figure*}
\centering
	\includegraphics[width=8.0cm, height=6.0cm]{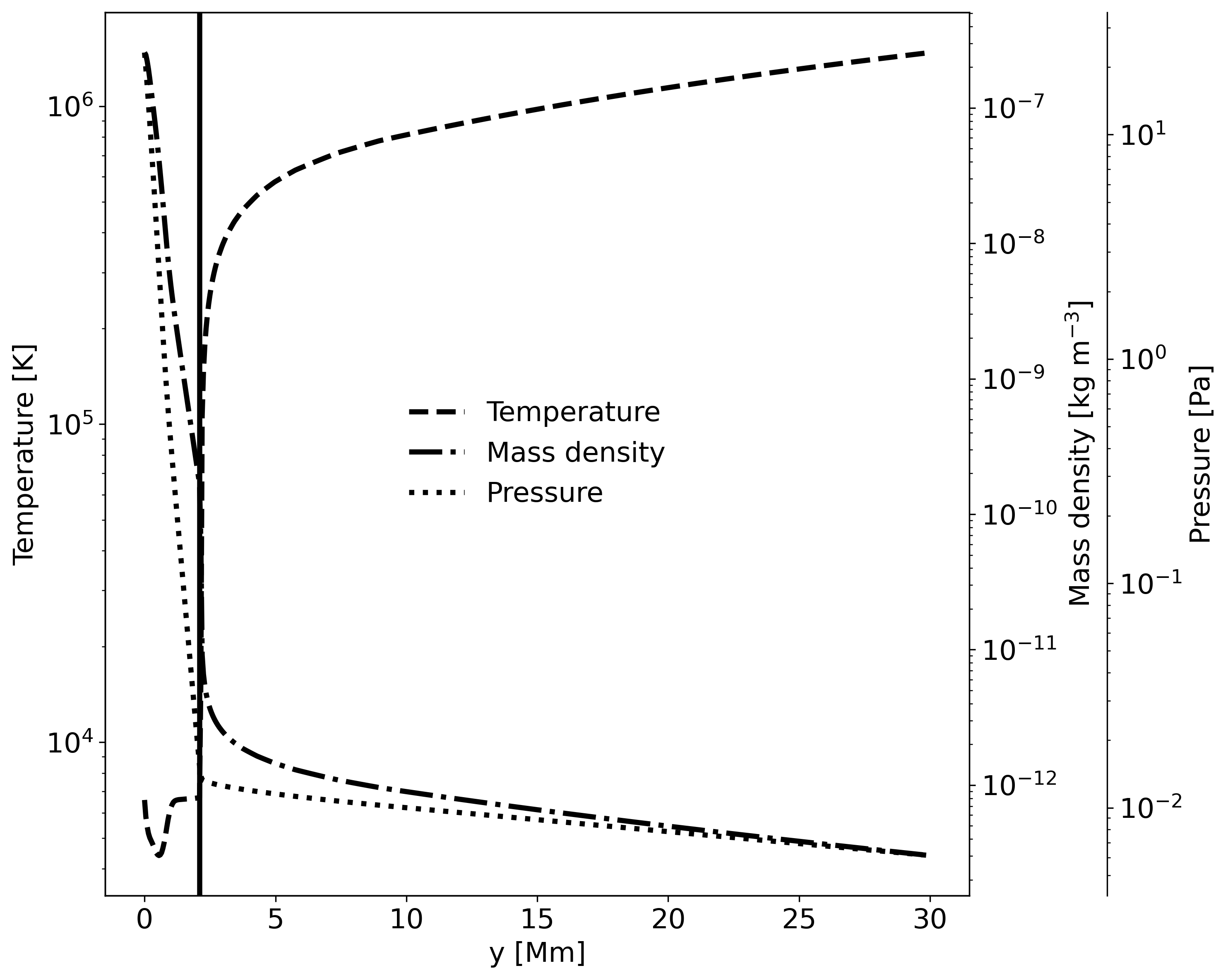} 
    \includegraphics[width=6.0cm, height=6.0cm]{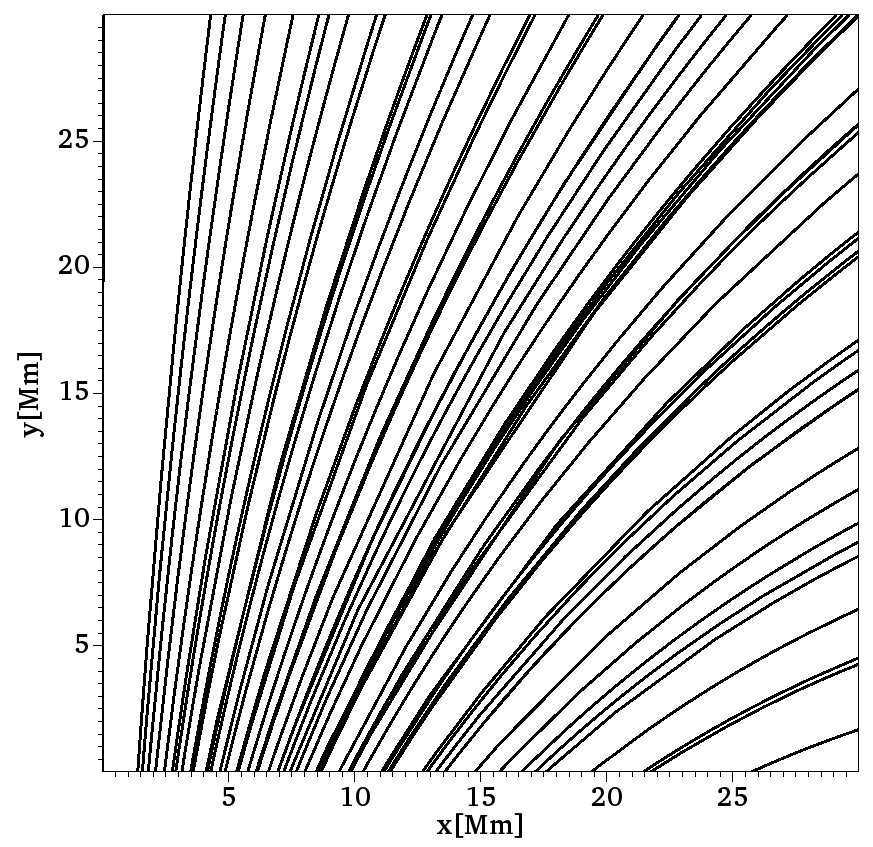}
    \caption{(Left) The logarithm of temperature in Kelvin (dashed line), the logarithm of mass density in kg m$^{-3}$ (dash dot line), and the logarithm of gas pressure in Pa (dotted line) versus height, $y$ in Mm, for the C7 equilibrium solar atmosphere model at the initial time ($t=0$ s) of the simulation. The vertical dashed line (black) represents the location of the transition region at about $y\sim 2.1$ Mm. (Right) Representation of the magnetic field lines in the 2D domain at $t=0$ s.}
     \label{fig:C7_model_mag_field_lines}
\end{figure*}

We assumed that the solar atmosphere was in hydrostatic equilibrium at the initial simulation time ($t=0$ s). In particular, we choose the semi-empirical C7 model to describe the temperature field that covers the chromosphere and transition region \cite{Avrett&Loeser2008}, smoothly extended to the solar corona as, for example, \cite{Gonzalez-Avilesetal2017}. In the left of Figure \ref{fig:C7_model_mag_field_lines}, we show the equilibrium temperature, mass density, and gas pressure and profiles as functions of height $y$, where the steep gradient is discernible in the transition region $y\sim2.1$ Mm, as indicated by a vertical black dashed line. 

\subsubsection{The magnetic field}
\label{subsubsec:Mag_field_model}

The magnetic field ${\bf B}$ at the initial time is a 2D configuration that satisfies the force-free $(\nabla\times(\nabla\times{\bf B})) = {\bf 0})$ and current-free $(\nabla\times{\bf B} = {\bf 0})$ conditions, with the following components: 

\begin{eqnarray}
 B_{x}(x,y) &=& \frac{-2S(x-a)(x-b)}{[(x-a)^{2}+(y-b)^{2}]^{2}}, \\
 B_{y}(x,y) &=& \frac{S(x-a)^{2}-(x-b)^{2}}{[(x-a)^{2}+(y-b)^{2}]^{2}},
\end{eqnarray}
where $S$ corresponds to the pole's magnetic field strength given in units of G Mm$^{2}$, and $a$ and $b$ are parameters that define the magnetic pole's location. In the right of Figure \ref{fig:C7_model_mag_field_lines}, we show the magnetic field lines, where it is discernible that the field diverged at around $(x,y)=(5,1.75)$ Mm. This inclined geometry of the magnetic field affects the jet's trajectory, which mainly follows the field lines, as will be shown later in this paper. This magnetic field configuration has already been used in numerical simulations of impulsively driven jets \cite{Gonzalez-Avilesetal2021}.     

\subsubsection{The radiation parameters}
\label{sub-sub-sec:Radiation_parameters}

In this paper, we consider a plasma described by the non-ideal MHD coupled with the radiation transport equations, for which we adopt a gray approximation following a two-moment approach by imposing the M1 closure. This closure describes radiative transport in diffusion and free-streaming limits and can describe highly anisotropic radiation transport \cite{Melon-Fuksmanetal2021}. 

For radiation parameters, we chose the scattering opacity, $\sigma = 0.34$ cm$^{2}$ g$^{-1}$, the absorption opacity, $\kappa = 1\times10^{-6}$ cm$^{2}$ g$^{-1}$. The value of the scattering opacity corresponds to the electron scattering opacity, which dominates the corona's fully ionized, optically thin plasma. This opacity is given by: 
\begin{equation}
\sigma = \frac{\sigma_{T}}{m_{p}},
\end{equation}
where $\sigma_{T}=6.65\times10^{-25}$ cm$^{2}$ is the Thomson scattering cross-section for free electrons, and $m_{p}=1.67\times10^{-24}$ g is the proton mass. Assuming a fully ionized hydrogen plasma (where the electron number density is approximately equal to the proton number density), the opacity per gram of plasma is $\sigma\approx 0.398$ cm$^{2}$ g$^{-1}$. However, helium is also present in small fractions in the solar corona, which slightly reduces the free-electron fraction per unit mass. When typical coronal abundances are included (e.g. $He/H\approx 0.1$), the effective opacity is reduced to approximately 0.34 cm$^{2}$ g$^{-1}$ \cite{1978stat.book.....M,2008oasp.book.....G}. For the value of the absorption opacity ($10^{-6}$ cm$^2$ g$^{-1}$), in the solar corona, it is much lower than the scattering opacity because the primary absorption mechanisms (bound-free and free-free absorption) are weak at coronal temperatures ($T\sim10^{6}$ K). The dominant source of absorption is free-free absorption (Bremsstrahlung), given by $\kappa\approx10^{-6}$ cm$^{2}$ g$^{-1}$ under typical coronal conditions \cite{1979rpa..book.....R,2000asqu.book.....C}.

At the initial time ($t=0$ s) of the simulation, we set the radiation energy, $E_{r} = 2.2$ g cm$^{-1}$ s$^{-2}$ and the radial component of the radiation flux, ${\bf F}_{r} = F_{rr} = 2.1$ g cm$^{-1}$ s$^{-2}$, which are typical values of the solar corona \cite{2020LRSP...17....3L}. Also, at the initial simulation time, the solar corona is not in local thermodynamic equilibrium (LTE), since the radiation energy density does not follow the equation $a_{R}T^{4}$, and the radiation flux is not equal to zero. If we consider a system in LTE, the radiation flux in the solar corona is high ($\sim 10^{16}$ J m$^{-3}$), which dominates the plasma total energy density and produces instabilities in the numerical solution of the radiation MHD equations. Besides, physically, the solar corona does not necessarily radiate as an isotropic black body, since anisotropic effects, such as thermal conduction and coronal heating, should be added. Furthermore, the solar corona is optically thin for most photospheric radiation, and only negligible absorption or scattering of radiative energy by the gas occurs \cite{2020LRSP...17....3L}. Even so, radiation scattering is important in the solar chromosphere. So, non-local thermodynamic equilibrium (NLTE) effects should be considered, since the ionization balance of hydrogen and helium is out of equilibrium, and assuming LTE to compute opacities is generally no longer accurate. In particular, the proper modeling of energy exchange must take into account the effects of NLTE, as implemented, for example, in \cite{Sukhorukov&Leenaarts2017, Gudiksenetal2011,Carlsson&Leenaarts2012}. 

We set the reduced value of the speed of light $\hat{c} = 0.1$, which has physical units of velocity and is part of the reduced speed of light approximation (RSLA). The need to use the RSLA is related to the ratio of the radiation propagation speed, $c$, to the maximum acoustic signal speed of the gas, which can be quite large in many astrophysical settings. Therefore, the ratio of the corresponding Courant-Friedrichs-Lewy (CFL) time steps for explicit integration of the gas and radiation transport subsystems, $\Delta t_{gas}/\Delta t_{rad}\sim c/v_{max}$, may be many orders of magnitude greater than 1. So an explicit scheme for the radiation subsystem can be rendered impractical by such a large ratio. Fortunately, in many situations we can reduce the signal propagation speed of the radiation fluid to some value $\hat{c}<<c$, which in turn reduces the explicit time step ratio of gas to radiation to a computationally tractable level, while preserving the essential dynamical behavior of the HD or MHD system. This is the essence of RSLA, initially described by \cite{GNEDIN2001437} and implemented by \cite{Gonzalezetal2007,10.1111/j.1365-2966.2008.13223.x,10.1111/j.1365-2966.2011.18986.x}. So in the case of this paper, we chose a value of $\hat{c}=0.1$ that preserves the dynamic behavior of the MHD system and makes the evolution of the radiation subsystem computationally tractable. We also tested the simulation with four other different values of $\hat{c}=10^{-2}, 10^{-4}, 10^{2}$ and $10^{4}$. Yet, the results do not vary significantly regarding the jet's formation and evolution. The main effect of a specific choice of $\hat{c}$ is reflected in the computational time. For example, for values $\hat{c}=10^{2}$ or $\hat{c}=10^{4}$, the computational time increases considerably, making the calculations impractical for the scenario presented in this paper. Therefore, $\hat{c}=10^{-1}$ is reasonable and appropriate in terms of computational cost. The choice of reduced speed of light, $\hat{c}$, was neither entirely arbitrary nor made solely for computational convenience. The rationale behind the Reduced Speed of Light Approximation (RSLA) is straightforward: since most astrophysical dynamics are modeled in the Newtonian limit (that is, the leading term in a Taylor expansion in powers of $1/c$), it is sufficient that the higher-order terms remain small compared to the leading one. For a system with characteristic velocity $v$, these terms scale as $v/c$. As long as $v/c \ll 1$, the Newtonian approximation holds. Therefore, the exact value of $c$ is not critical, provided that $v \ll \hat{c}$, where $\hat{c}$ is the reduced speed of light, typically chosen as a fraction of the true speed of light (e.g., $\hat{c} = 0.1c$, $0.5c$, or $0.7c$). In this study, the problem lies well within the Newtonian regime, as the characteristic velocity of the jet is on the order of tens of km~s$^{-1}$, which is much lower than the adopted value of $\hat{c} = 0.1c$. This justifies the use of a reduced speed of light, which does not affect the physical dynamics but significantly reduces the computational cost. Thus, $\hat{c} = 0.1c$ is an appropriate and efficient choice for the simulations presented in this work.

Therefore, in this paper, the time-dependent formulation of radiative transfer has the following advantages: i) considering the evolution of the plasma variables coupled with the radiation energy density and radiation flux, and ii) adopting a gray approximation following a two-moment approach by imposing the M1 closure, which covers radiative transport in diffusion and free-streaming limits and can describe highly anisotropic radiation transport. However, the main disadvantages are: i) neglecting the effects of radiation scattering in the solar chromosphere and the NLTE effects, and ii) simplification of the treatment of the radiation field by considering the frequency-integrated zeroth- and first-order angular moments of the radiative transfer equation.

 \subsubsection{Perturbation}
\label{sub-sub-sec:Perturbation}

To excite jet formation, we perturb the initial background state of the solar atmosphere by a simple localized Gaussian pulse set in the vertical component of velocity, which is a standard procedure for triggering jets \cite{Muraswki&Zaqarashvili2010, Muraswkietal2011, Gonzalez-Avilesetal2021}. Specifically, the velocity pulse is  

\begin{equation}
    v_{y}(x,y) = A_{v}\exp\left(-\frac{(x-x_{0})^{2}+(y-y_{0})^{2}}{w^{2}}\right).
\end{equation}
In the above expression, $A_{v}$ defines the amplitude of the pulse, $(x_{0},y_{0})$ is the initial position of the pulse, and $w$ is its width. In this paper, we set and hold $A_{v} = 100$ km s$^{-1}$, $x_{0} = 5$ Mm, $y_{0} = 1.75$ Mm, and $w=0.2$ Mm. These values are typical for exciting jets that resemble some macrospicule characteristics in adiabatic and non-adiabatic numerical MHD simulations \cite{Muraswkietal2011, Gonzalez-Avilesetal2021}. 

\subsection{Numerical methods}
\label{sub-sec:Num_methods}

We numerically solve equations (\ref{mass_cons_eq})–(\ref{flux_rad_eq}) using version 4.4-patch3 of the PLUTO code \cite{Mignoneetal2007}, which incorporates a radiation module based on the two-moment radiation hydrodynamic scheme described in \cite{Melon-Fuksmanetal2021}. The simulations employ a Courant-Friedrichs-Lewy (CFL) number of 0.4 and use a second-order total variation-diminishing (TVD) Runge-Kutta time integrator. For the integration of the MHD equations, we adopt the approximate Harten-Lax-van Leer discontinuities (HLLD) Riemann solver \cite{MIYOSHI2005315}, while the radiation transport equations are solved using the Harten-Lax-van Leer contact solver (HLLC) \cite{LI2005344}. In both cases, a linear reconstruction of the primitive variables is applied, combined with the minmod slope limiter. 

To control the numerical violation of the divergence-free condition ($\nabla\cdot\mathbf{B} = 0$), we adopt the constrained transport method (CT), which computes the electric field $\mathbf{E}$ using a simple arithmetic averaging scheme \cite{BALSARA1999270}. This approach ensures that the solenoidal constraint on the magnetic field is preserved to machine precision throughout the numerical evolution. The integration of the radiation transport step follows the methods implemented in \cite{Melon-Fuksman2019}. Specifically, equations (\ref{energy_rad_eq}) and (\ref{flux_rad_eq}) are advanced in time using Implicit-Explicit Runge-Kutta (IMEX-RK) schemes \cite{ascher1997imex}. In these schemes, flux terms are treated explicitly, while radiation–matter interaction terms are implicitly integrated. PLUTO provides two IMEX schemes for this purpose: the IMEX-SSP2(2,2,2) method \cite{Pareschi&Russo2005} and the IMEX1 scheme introduced by \cite{Melon-Fuksman2019}. The IMEX-SSP2(2,2,2) method is a two-stage second-order scheme with strong stability preserving properties (SSP), widely used for stiff hyperbolic-parabolic systems. In contrast, the IMEX1 scheme is a first-order formulation tailored for MHD with parabolic terms, designed to efficiently handle stiff dissipative terms while maintaining simplicity and robustness. Thermal conduction is evolved separately from the advection terms via operator splitting, using the super-time-stepping (STS) method \cite{Alexiadesetal1996}. This approach is particularly well-suited for efficiently handling the highly parabolic nature of the thermal conduction term in the MHD equations. Lastly, the empirical coronal heating term in the energy equation (\ref{energy_cons_eq}) is explicitly integrated in time, without imposing additional constraints on the time step.

The simulation is carried out in a two-dimensional Cartesian domain with spatial coordinates $x \in [0, 30]$ Mm and $y \in [0, 30]$ Mm, discretized on a uniform grid of $1200 \times 1200$ cells. This corresponds to an effective spatial resolution of 25 km in both directions. In this setup, the lower boundary at $y = 0$ Mm corresponds to the base of the photosphere. For magnetohydrodynamic (MHD) variables, we impose fixed-time boundary conditions, maintaining all quantities at their equilibrium values along the edges of the domain at $x = 0$, $x = 30$ Mm, $y = 0$, and $y = 30$ Mm. In contrast, for radiation variables, namely the radiation energy density $E_{r}$ and the radiation flux vector $\mathbf{F}_{r}$, we apply the outflow boundary conditions at the four domain boundaries. These conditions allow radiation to freely exit the computational domain, ensuring a physically consistent treatment of radiative transport across the boundaries.

\section{Results of numerical simulations}
\label{sec:results_num_simulations}

\begin{figure*}
\centering
	\includegraphics[width=4.0cm, height=5.5cm]{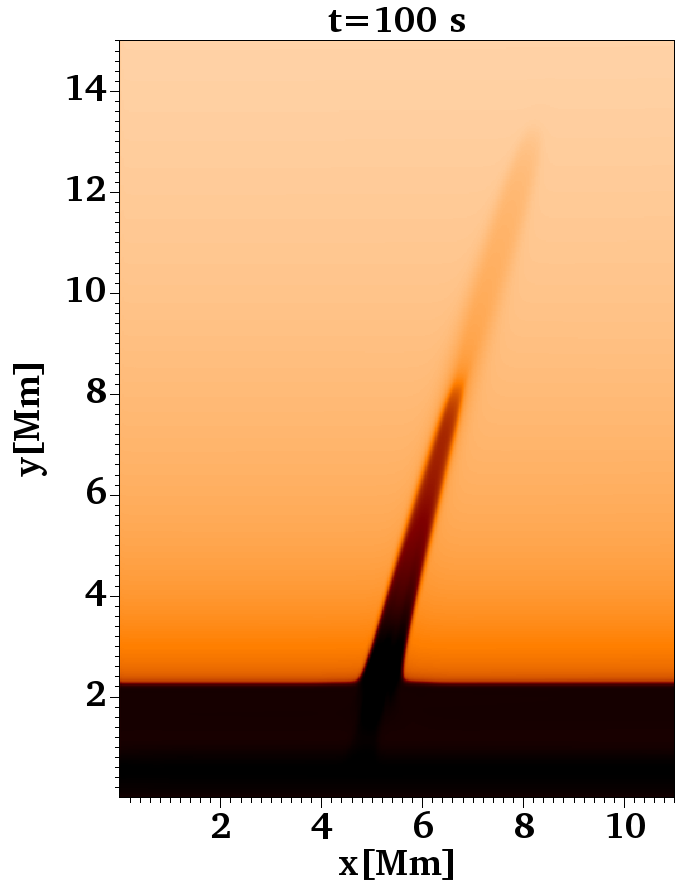}
    \includegraphics[width=4.0cm, height=5.5cm]{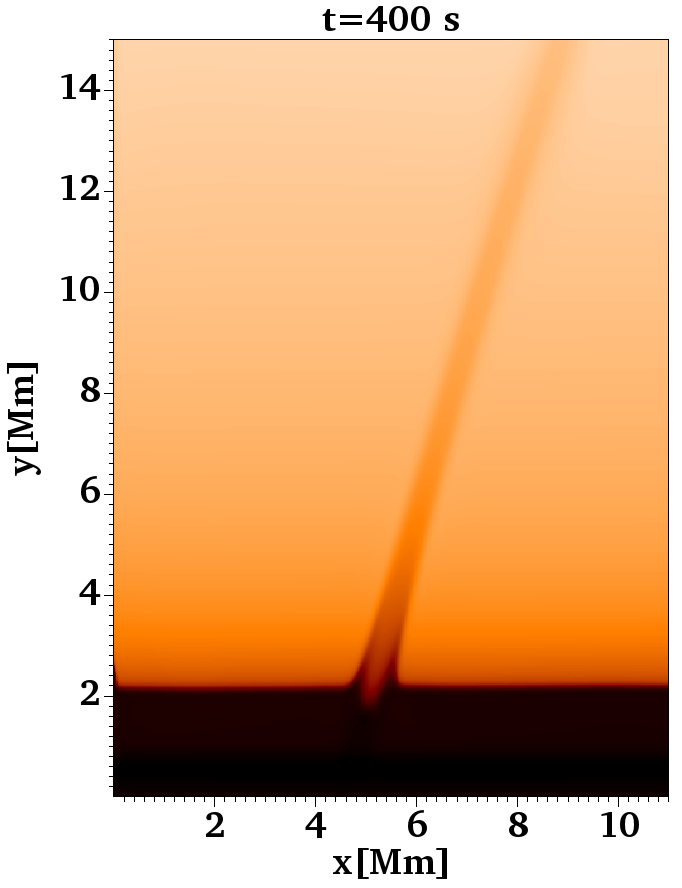}
    \includegraphics[width=5.0cm, height=5.5cm]{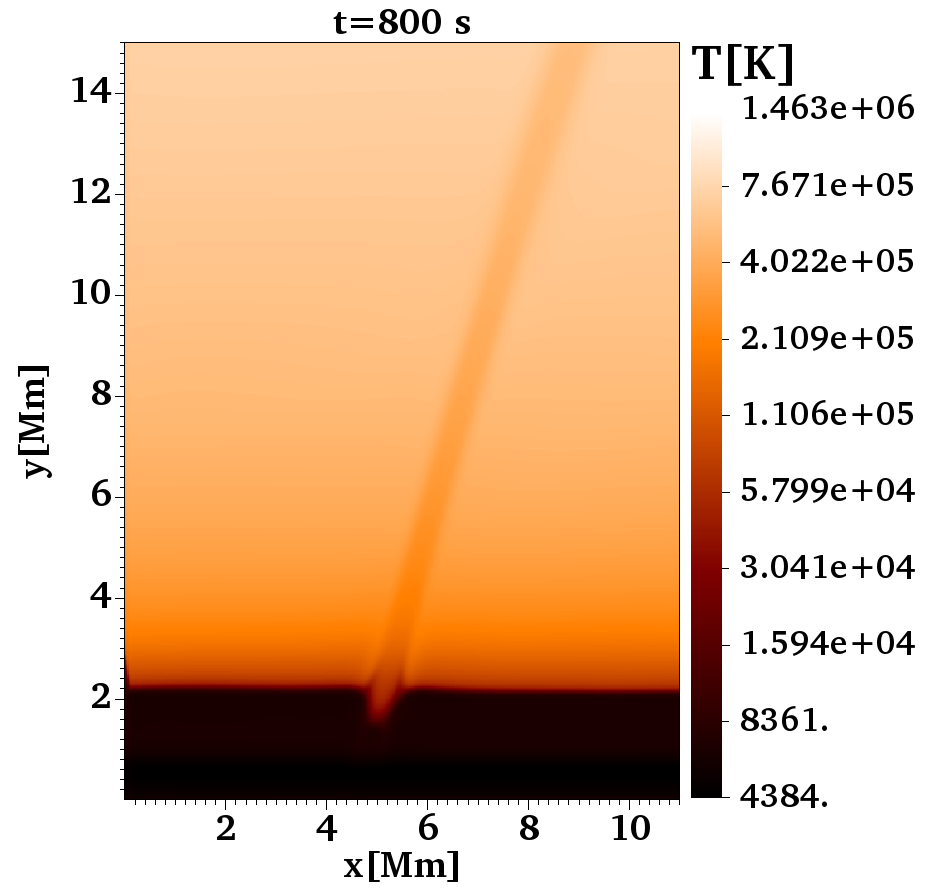}\\
    \includegraphics[width=4.0cm, height=5.5cm]{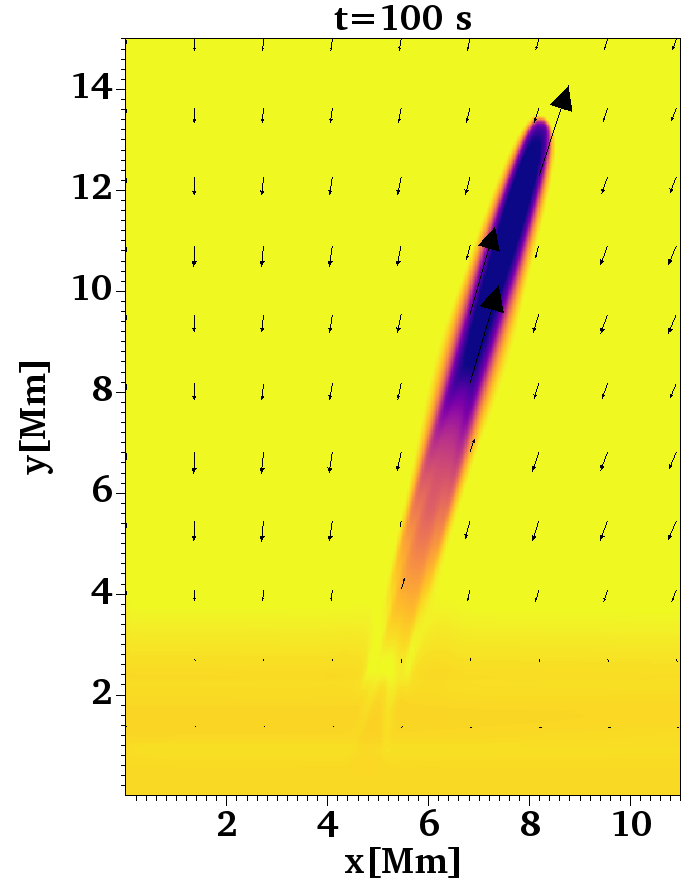}
    \includegraphics[width=4.0cm, height=5.5cm]{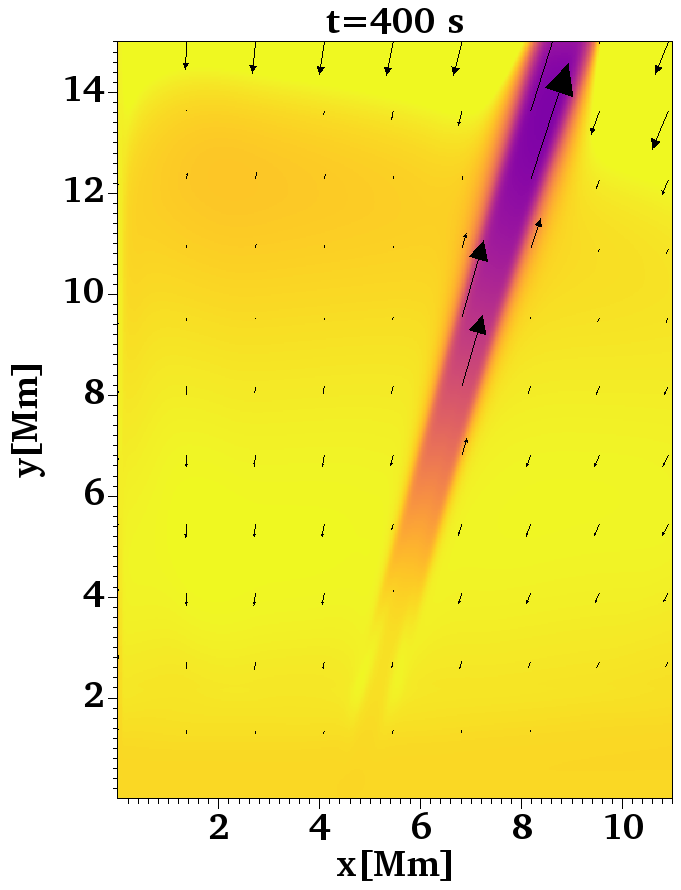}
    \includegraphics[width=5.0cm, height=5.5cm]{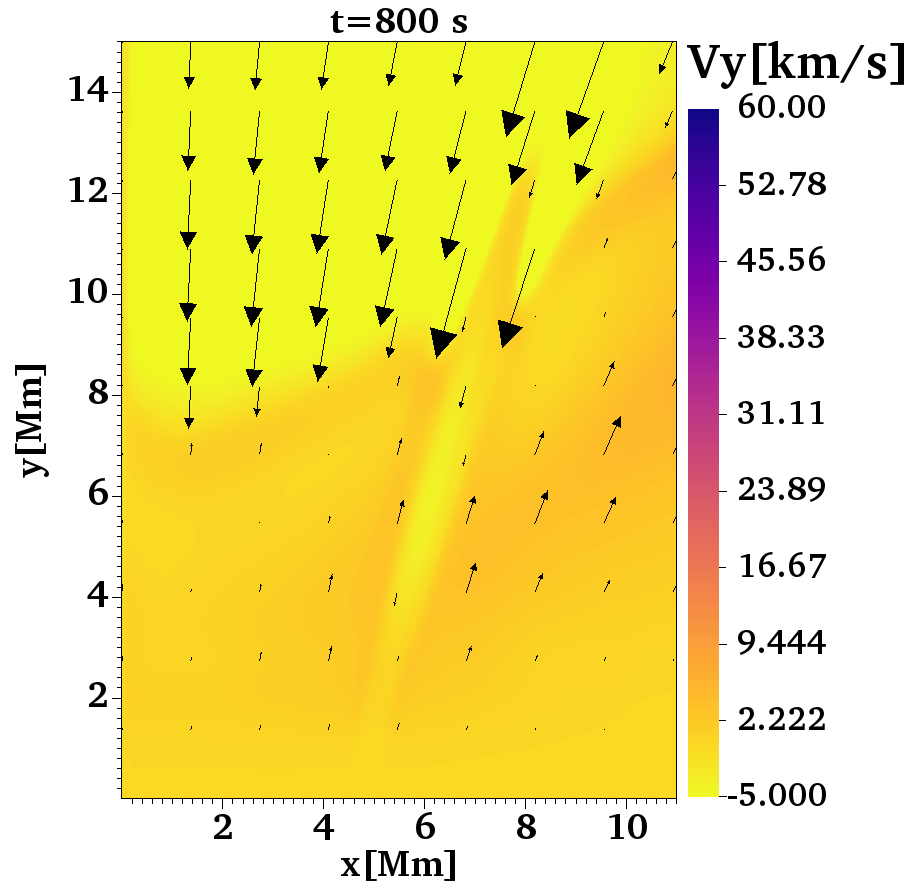}
    \caption{Top row: Spatial profiles of temperature (in Kelvin) at $t=100$ s (left), $t=400$ s (center), and $t=800$ s (left). Bottom row: Vertical component of velocity $V_{y}$ in km s$^{-1}$ with velocity vector field at the exact three times as temperature.}
    \label{fig:Temp_Vy_evolution}
\end{figure*}

In Figure \ref{fig:Temp_Vy_evolution}, we show snapshots of the spatial profiles of plasma temperature in Kelvin in the top panels and the vertical component of the velocity in km s$^{-1}$ with the vector velocity field overplotted in the bottom panels. Specifically, on the temperature map at $t=100$ s, we observe the formation of a collimated plasma structure that resembles a jet that reaches a height of about 8 Mm. This jet reaches a maximum temperature of approximately $8\times10^{4}$ K and also shows at its head the development of a collimated hot plasma structure ($\sim 2\times10^{5}$ K) that propagates upward. At $t=400$ s, the jet becomes smaller ($\sim 1.5$ Mm in length) compared to its precedent at $t=100$ s; it also exhibits a double-thread structure at the top of it, which is a typical feature of solar spicules \cite{1974IAUS...56..239T, 1998ESASP.421..345V}. At $t=800$ s, the temperature map shows that the jet practically diffuses and that only its tiny double thread remains over the transition region ($\sim 2.1$ Mm) and the upper chromosphere ($\sim 3$ Mm). In the bottom row of Figure \ref{fig:Temp_Vy_evolution}, we display vertical velocity maps $V_{y}$ with the vector velocity field showing consistency with the temperature behavior described above. In particular, at $t=100$ s, the jet reaches a vertical velocity of about 30 km s$^{-1}$, while the plasma at the top of it reaches a maximum high vertical velocity of around 60 km s$^{-1}$). This high-velocity is associated with a shock front generated by the perturbation that moves from a cooler (denser) medium to a hotter (less dense) medium into the solar corona. This is a typical behavior obtained in numerical simulations of impulsively driven jets \cite{Muraswki&Zaqarashvili2010}.
However, at $t=400$ s, the plasma associated with the shock reaches the top boundary at $y=30$ Mm. At $t=800$ s, the vertical velocity is predominantly negative, indicating the jet diffusion. Interestingly, radiative terms in the momentum and total energy density equations and their coupling with the plasma could generate this negative plasma-flow motion. This result, in turn, is consistent with the effect of the optically thin radiative losses, as shown in \cite{Gonzalez-Avilesetal2021}. 

\begin{figure*}
    \centering
    \includegraphics[width=6.5cm, height=4.5cm]{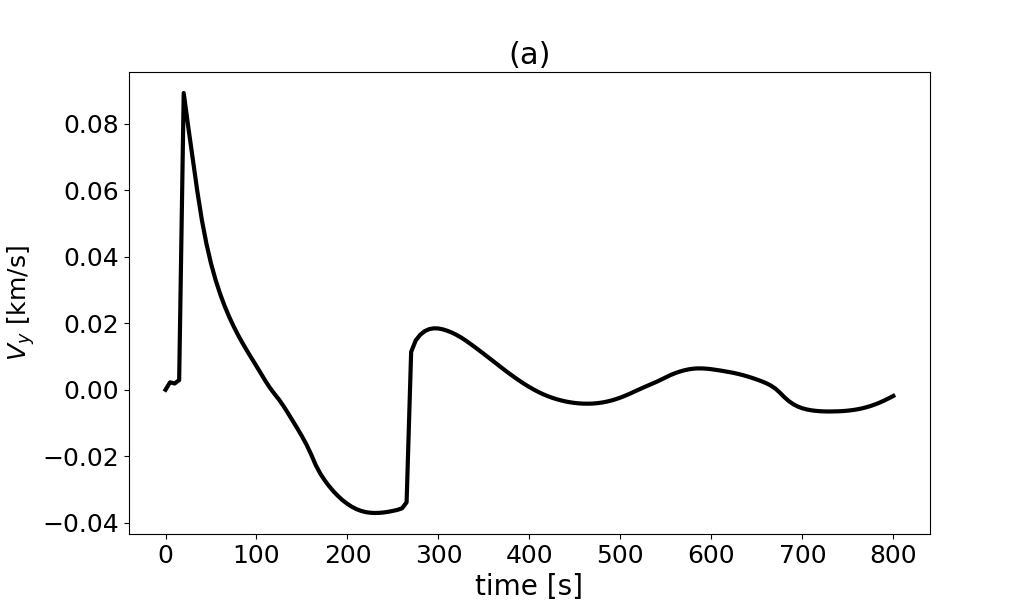}
    \includegraphics[width=6.5cm, height=4.5cm]{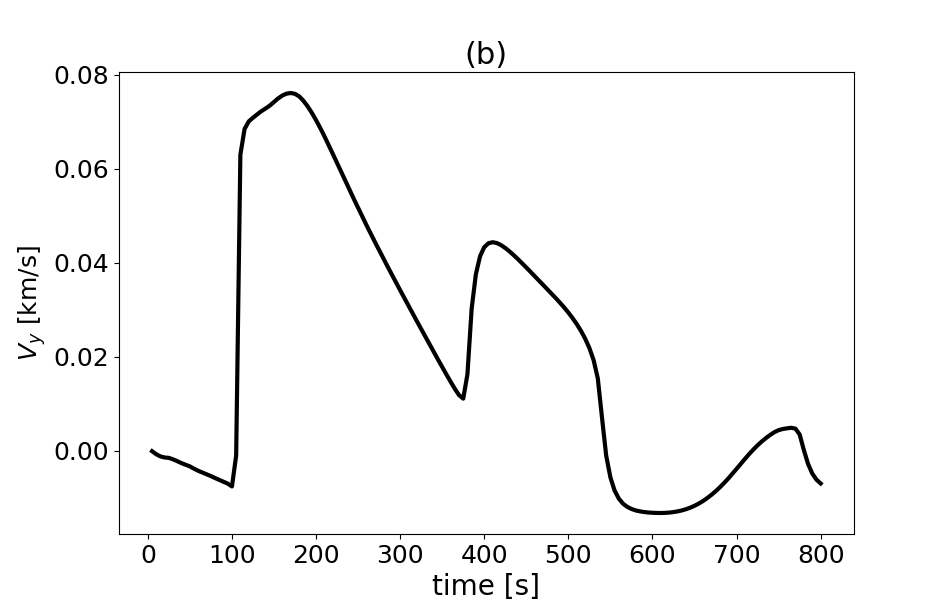}
     \includegraphics[width=6.5cm, height=4.5cm]{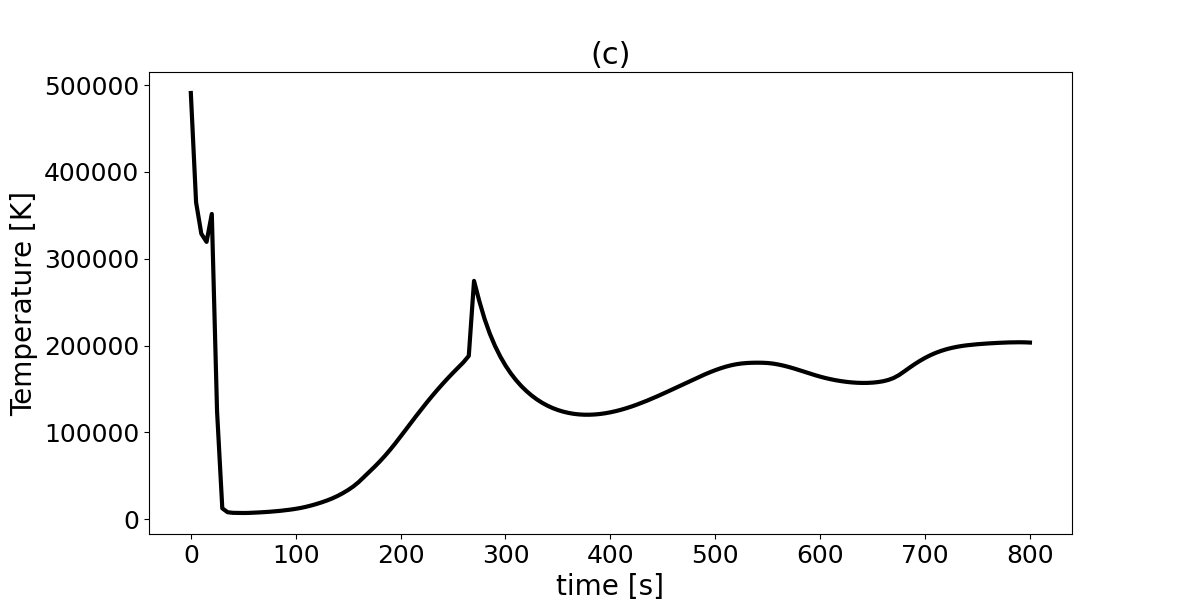} 
     \includegraphics[width=6.5cm, height=4.5cm]{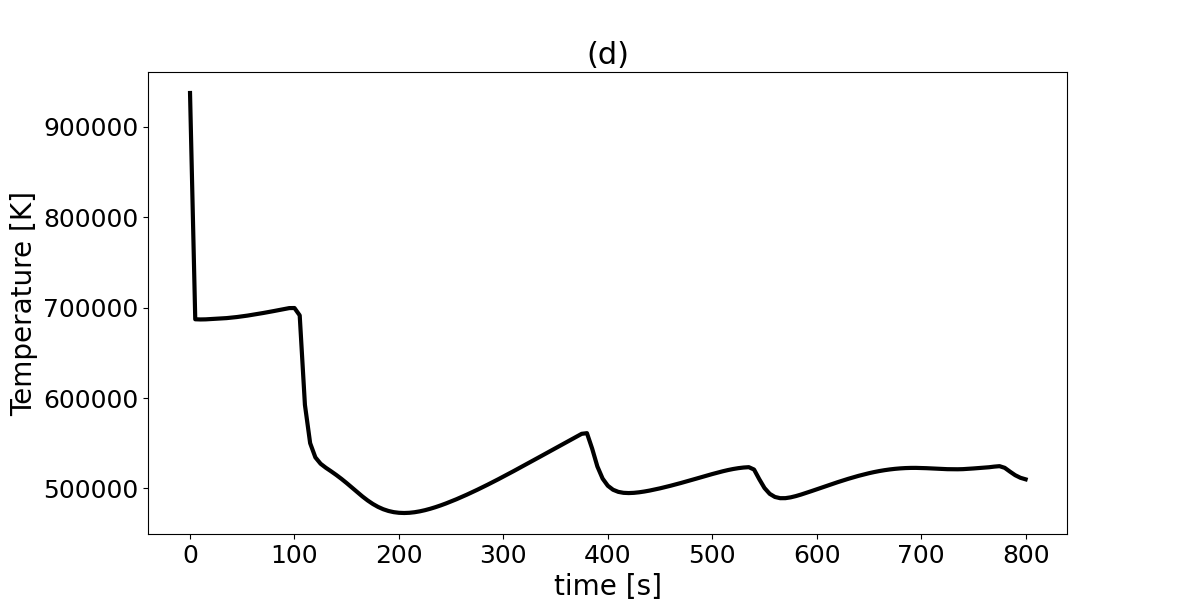} \\
     \includegraphics[width=6.5cm, height=4.5cm]{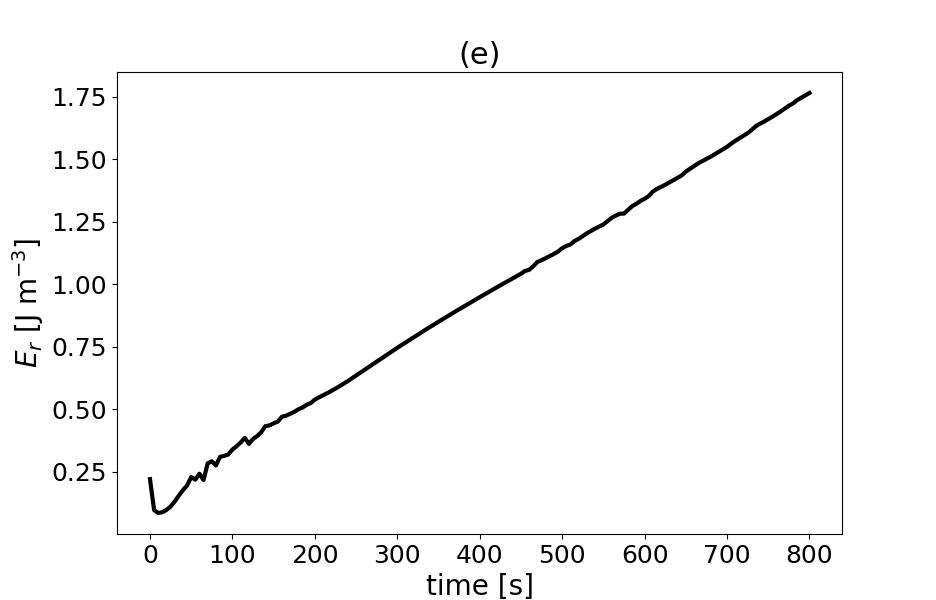}
      \includegraphics[width=6.5cm, height=4.5cm]{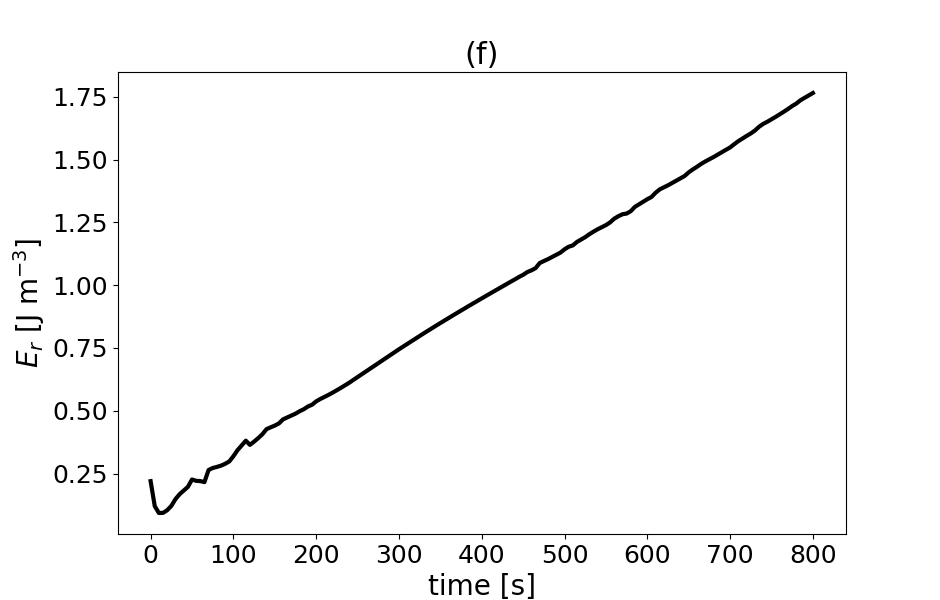}\\
     \includegraphics[width=6.5cm, height=4.5cm]{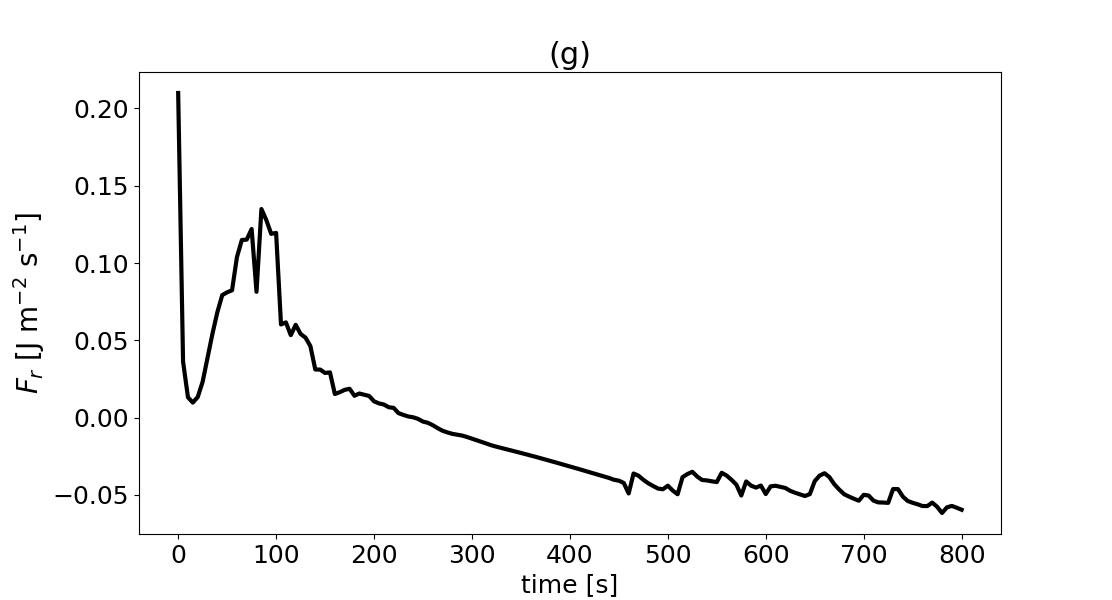}
     \includegraphics[width=6.5cm, height=4.5cm]{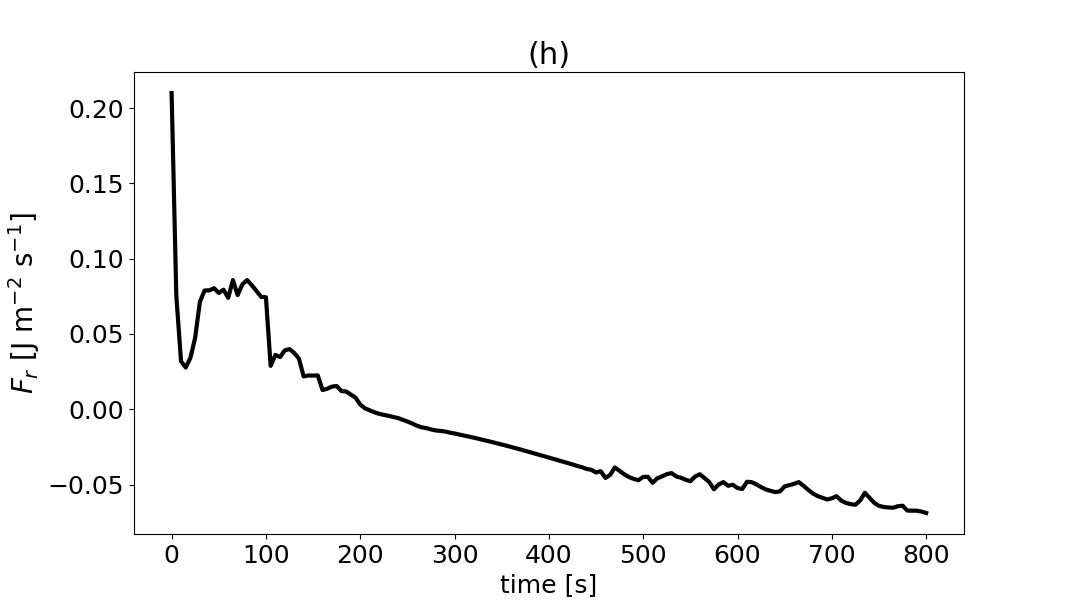}
    \caption{Time signatures of the vertical velocity, $v_{y}$, in km s$^{-1}$ (a)-(b), the temperature, in Kelvin (c)-(d), the radiation energy density, $E_{r}$, in J m$^{-3}$ (e)-(f), and the radial component of the radiation flux, $F_{r}$, in J m$^{-2}$ s$^{-1}$ (g)-(h), collected at the point $x=5.6, y = 4$ Mm (left panels) and $x=8.5, y = 14$ Mm (right panels).}
    \label{fig:time_series}
\end{figure*}

In Figure \ref{fig:time_series}, we show the time signatures of the vertical velocity $v_{y}$ (km s$^{-1}$), temperature (K), radiation energy density $E_{\mathrm{nr}}$ (J m$^{-3}$) and the radial component of the radiation flux $F_{r}$ (J m$^{-2}$ s$^{-1}$) at the detection points $(x = 5.6, y = 4)$~Mm (left panels) and $(x = 8.5, y = 14)$~Mm (right panels). In this paper, a time signature refers to the temporal evolution of a physical quantity at a fixed spatial location, used to identify the characteristic temporal behavior or periodicity in the time series of the plasma and radiation variables described above. In these time signatures, the point $(x=5.6, y=4)$ Mm lies within the jet and $(x=8.5, y=14)$ Mm is where the shock passes. In panel (a) of Figure~\ref{fig:time_series}, we show the time signatures of $v_{y}$ at $(x=5.6, y=4)$ Mm, where the vertical velocity inside the jet reaches a maximum value of about 90 km s$^{-1}$ around 20\,s after the simulation starts. After this stage, the vertical velocity becomes negative up to around -40 km s$^{-1}$ at $t=250$ \,s. Following this behavior, the velocity increases again, with values of 20 km s$^{-1}$. From around $t=350$ s onward, the vertical velocity within the jet remains negative, indicating diffusion. In panel (b) of Figure~\ref{fig:time_series}, the time signatures of the vertical velocity collected at the point $(x=8.5, y=14)$ Mm indicate that the shock moves fast and reaches a maximum velocity of around 80 km s$^{-1}$ at approximately $t=180$\,s. After that time, the vertical velocity of the shock decreases to a value near 18 km s$^{-1}$ at about $t=360$\,s. Later, the vertical velocity increases again to a value of around 45 km s$^{-1}$. Nonetheless, after 400\,s, the vertical velocity of the shock decreases and reaches negative values. This behavior indicates that plasma inside the shock moves downwards from the solar corona to the transition region and chromosphere.       
In panel (c) of Figure~\ref{fig:time_series}, the time signatures of temperature measured at the detection point $(x=5.6, y=4)$ Mm show that the temperature inside the jet rapidly decreases at the start of the simulation ($t\approx 50$ s). Later, the temperature increased to an order of magnitude of $10^{5}$ K close to 300 s. From 300 to 800 s, the temperature varies slightly and remains close to $2\times10^{5}$ K. In panel (d), the time signatures of the temperature collected at $(x=8.5, y=14)$ Mm indicate that the temperature inside the shock is in the range of $7\times10^{5}-9\times10^{5}$ K in the first $100$\,s of simulation. After that time, the temperature of the shock decreases to a value close to $3\times10^{5}$ K, at $t=200$ s. After 200 s, the temperature slightly varies around $5\times10^{5}$ K to reach the final simulation time $t=800$ s.   

In panels (e) and (f) of Figure~\ref{fig:time_series}, we show the time signatures of the radiation energy density collected at two detection points described above. In these time histories, we identify a linear growth of the radiation energy density inside the jet and within the shock wave during the total simulation time. In particular, the radiation energy density reaches a maximum value of around 1.75 J m$^{-3}$ at both detection points. Finally, in panels (g) and (h), that is, at the bottom of Figure~\ref{fig:time_series}, the time signatures of the radial component of the radiation flux show the opposite behavior to the radiation energy density. Remarkably, the radiation flux decreases; this behavior could be associated with the radiation energy loss processes at the solar corona, which, in turn, might be related to an optically thin radiative cooling process.

\begin{figure*}
    \centering
    \includegraphics[width=8.0cm, height=5.0cm]{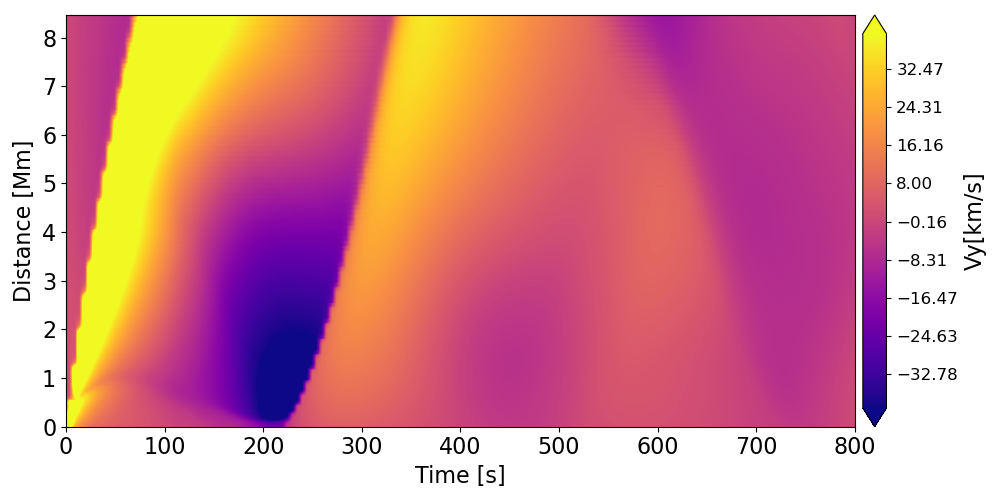}
     \includegraphics[width=8.0cm, height=5.0cm]{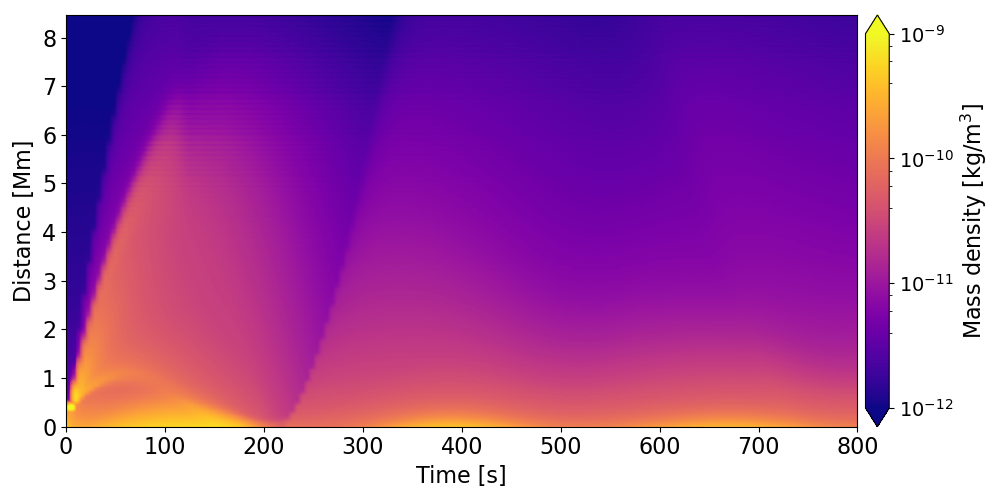}
    \caption{Distance-time diagrams of the vertical velocity $v_{y}$, in km s$^{-1}$ (left), and the logarithm of mass density, $\rho$, in kg m$^{-3}$ (right).}
    \label{fig:time-distance_diagram}
\end{figure*}

To identify the dynamics of the plasma associated with the jet and shock, in Figure~\ref{fig:time-distance_diagram}, we show distance-time diagrams of the vertical velocity $v_{y}$ in km s$^{-1}$ and the logarithm of the mass density $\rho$ in kg m$^{-3}$ measured in a line of about 8.15 Mm in length that goes along with the jet, similar to the analysis performed in \cite{Gonzalez-Avilesetal2021}. For example, on the left of Figure~\ref{fig:time-distance_diagram}, the temporal and spatial evolution of the vertical velocity along the jet schematizes the generation of a shock wave combined with a contact wave at the beginning of the jet formation. In the right panel of Figure~\ref{fig:time-distance_diagram}, we depict the behavior of the mass density along the jet, which shows that the jet reaches a height of about 8 Mm at $t\approx 100$ s. It is also evident that there is periodic behavior in the jet dynamic, and it follows a typical parabolic path with a period of around 300 s, which could be related to a chromospheric jet such as spicules \cite{Hansteenetal2006}.

In summary, the numerical results described in the previous paragraphs reveal that the impulsively driven chromospheric jet rapidly develops into a collimated plasma structure reaching a height of about $8$~Mm, with temperatures up to $8\times10^{4}$~K and a hotter leading edge of $\sim 2\times10^{5}$~K. At late simulation times, the jet exhibits a characteristic double–thread morphology and eventually dissipates back into the lower atmosphere. In addition, vertical velocity attains values of $\sim 30$~km~s$^{-1}$ within the jet and up to $\sim 60$~km~s$^{-1}$ at the associated shock front, with time signatures at fixed locations indicating peak velocities close to $90$~km~s$^{-1}$ before declining and becoming negative as the structure decays. The temperature inside the jet stabilizes around $2\times10^{5}$~K, while in the shock region it decreases from $9\times10^{5}$~ K to around $3\times10^{5}$~K during simulation. Likewise, the radiation energy density increases linearly to $\sim 1.75$~J~m$^{-3}$, whereas the radiation flux decreases, consistent with optically thin radiative cooling. The distance-time diagrams confirm a parabolic trajectory with a period of $\sim 300$~ s, reproducing the classical spicule-like dynamics prior to the final dissipation of the jet.

Overall, the results of this section demonstrate that although the simulated jet reproduces some characteristics similar to classical spicules, its rapid fading is strongly influenced by radiative processes. In particular, the explicit evolution of the radiation energy density and flux through the two–moment transport equations reveals that radiation acts as an efficient optically thin cooling mechanism in the corona. This mechanism induces downward flows and thermal energy loss, thereby limiting the height, duration, and energy content of the jet. Consequently, the present findings not only agree with previous studies \citep{Sterling&Mariska1990, Gonzalez-Avilesetal2021} employing empirical radiative loss functions, but also emphasize that a self–consistent treatment of radiation transport is crucial to capture the dissipation and evolution of macrospicule–like events.

\section{Discussion}
\label{sec:discussion}

\subsection{Effects of the radiation transport terms on the jet}
\label{sub_sec:effects_rad_terms}

\begin{figure*}
\centering
	\includegraphics[width=4.0cm, height=5.0cm]{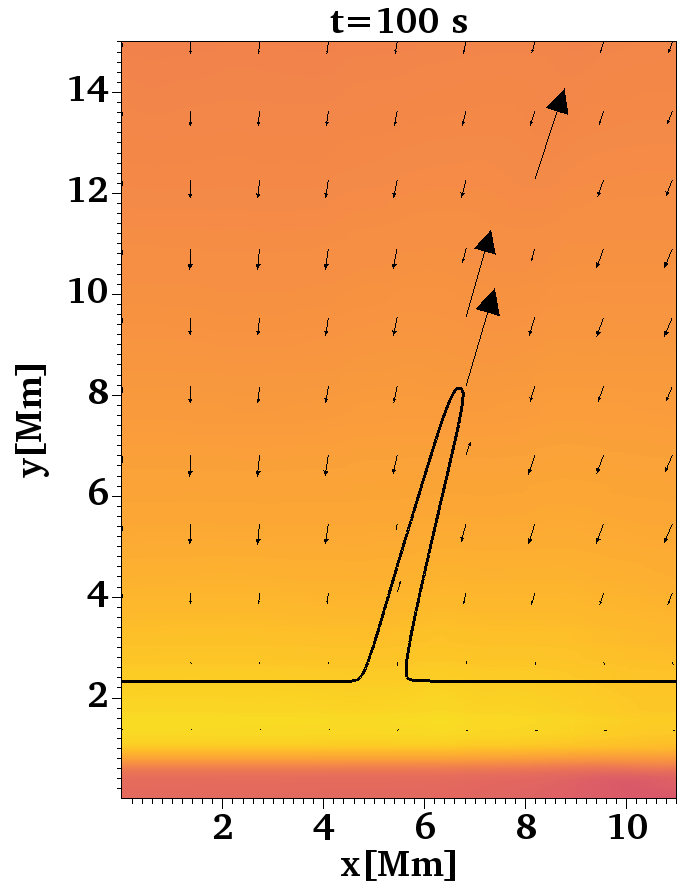}
    \includegraphics[width=4.0cm, height=5.0cm]{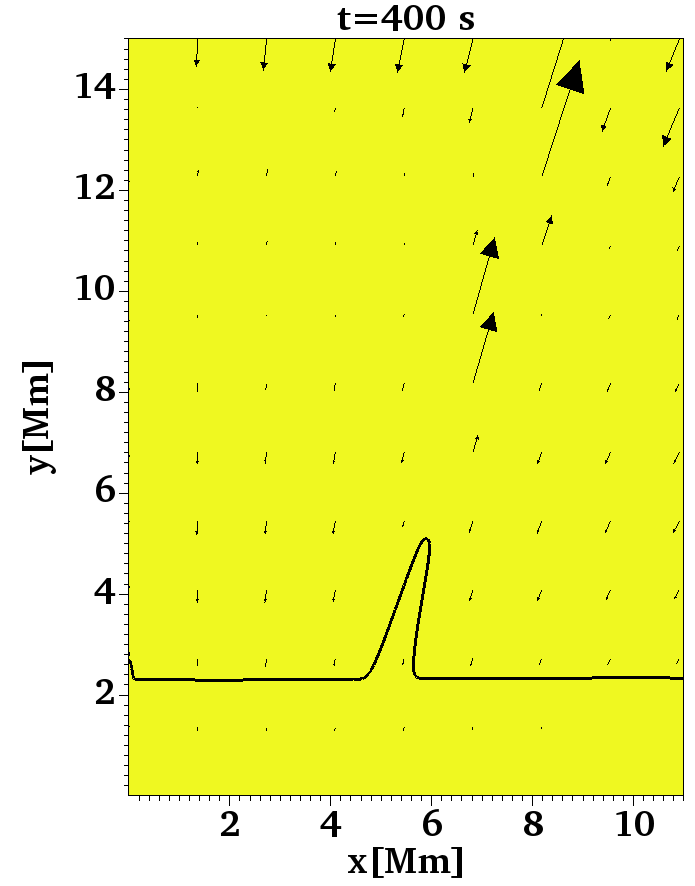}
    \includegraphics[width=5.5cm, height=5.0cm]{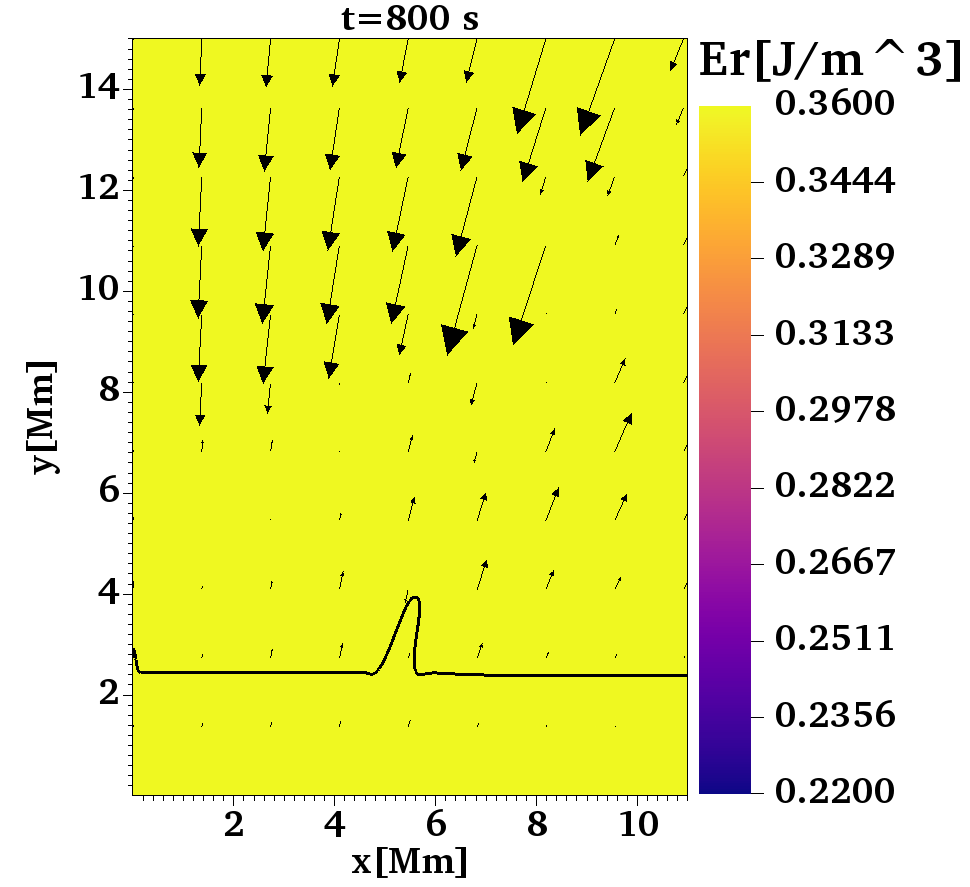}\\
    \includegraphics[width=4.0cm, height=5.0cm]{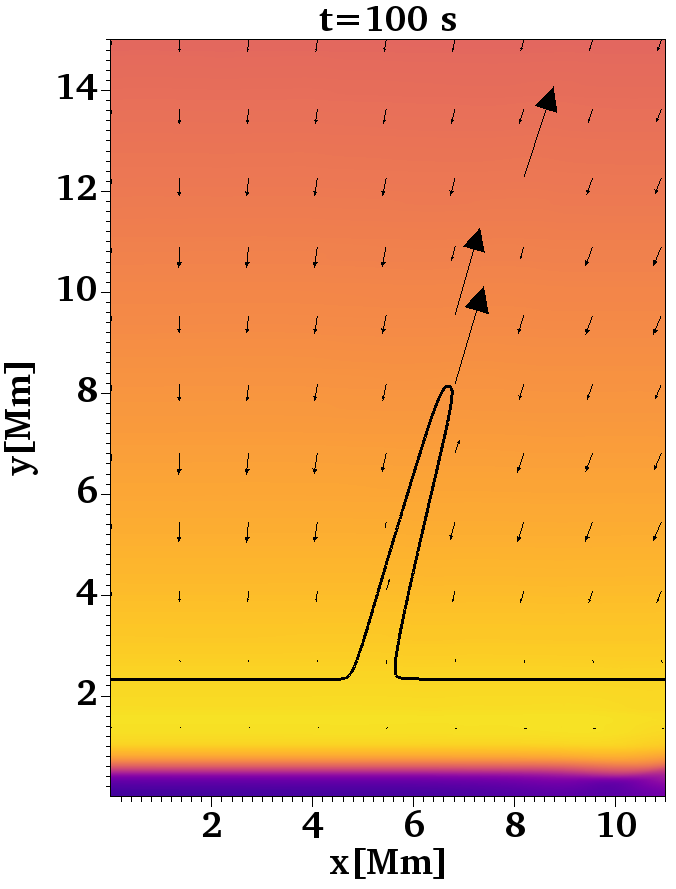}
    \includegraphics[width=4.0cm, height=5.0cm]{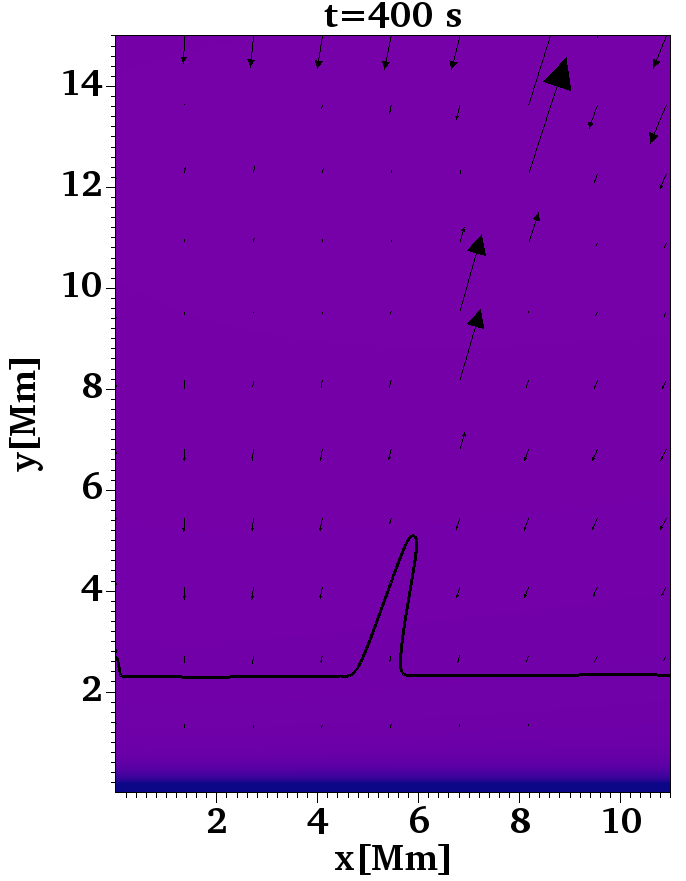}
    \includegraphics[width=5.3cm, height=5.0cm]{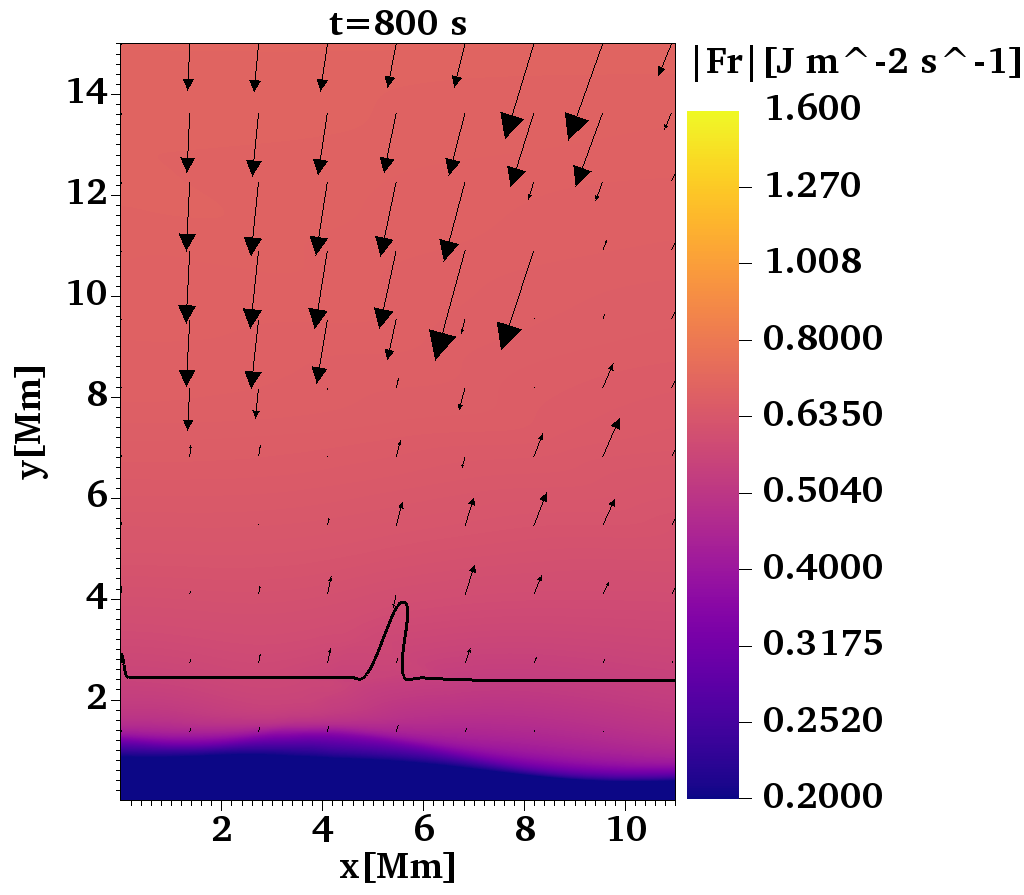}\\
    \includegraphics[width=4.0cm, height=5.0cm]{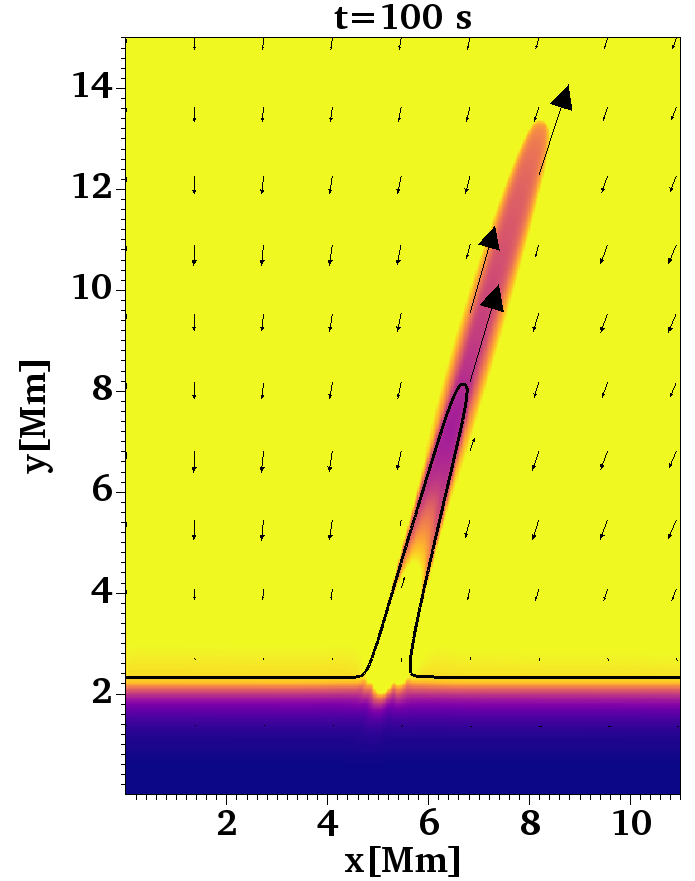}
    \includegraphics[width=4.0cm, height=5.0cm]{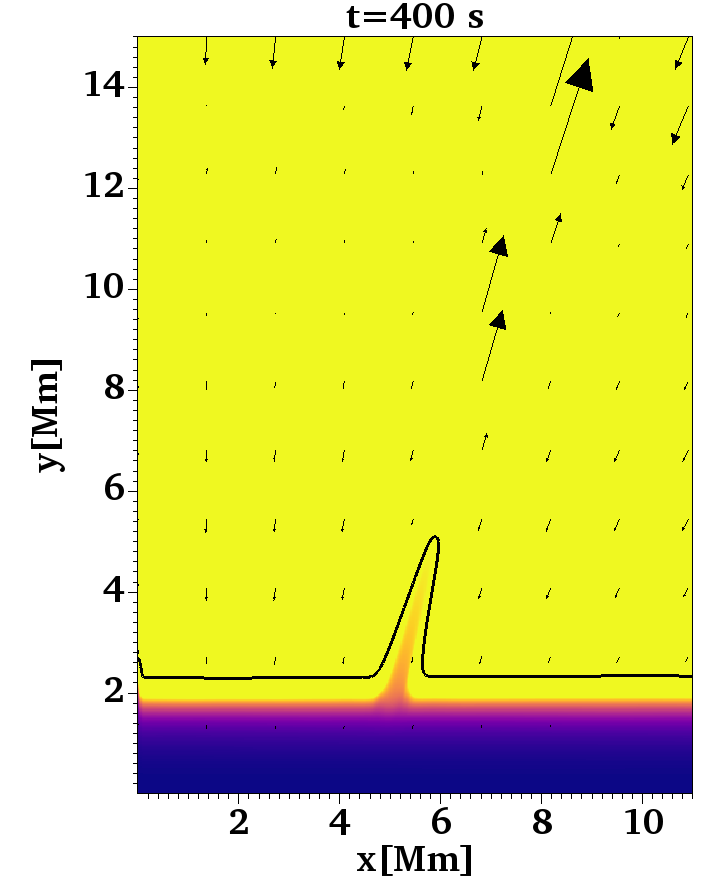}
    \includegraphics[width=5.5cm, height=5.0cm]{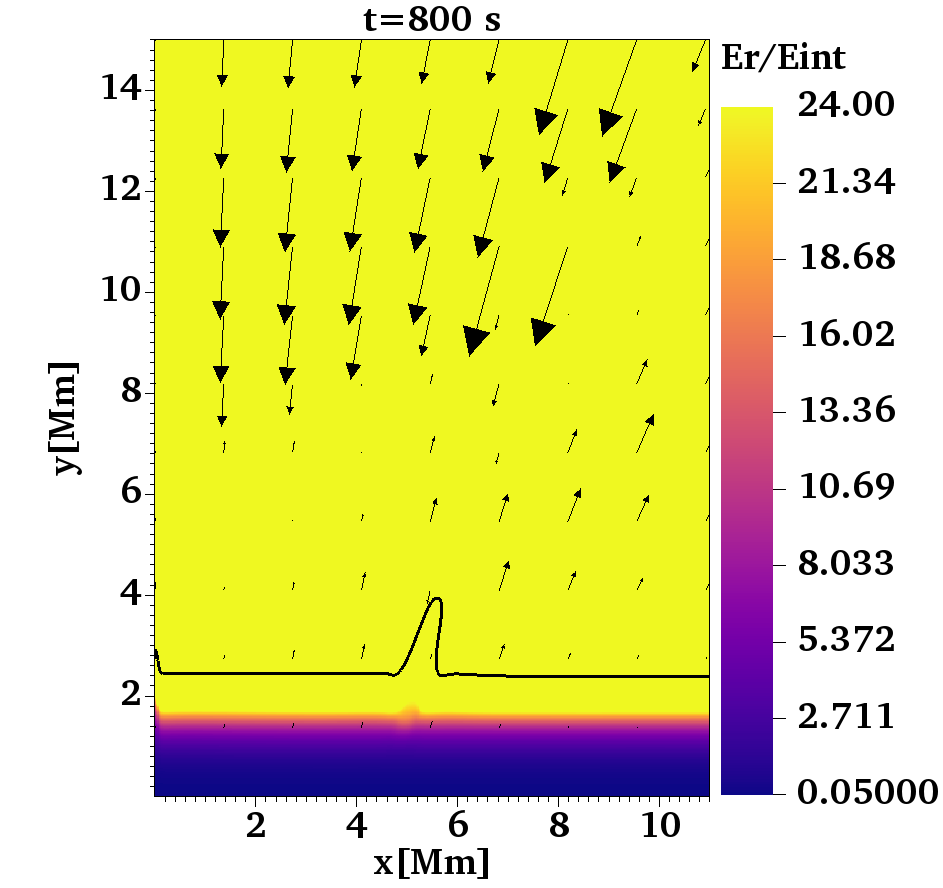}
    \caption{(Top) Spatial profiles of radiation energy density $E_{r}$ (in J m$^{-3}$), magnitude of the radiation flux $|{\bf F_{r}}|$ in J m$^{-2}$ s$^{-1}$ (middle), and the ratio $E_{r}/e_{int}$ (bottom), at the times $t=100$ s (left), $t=400$ s (center), and $t=800$ s (right). The vectors (in black) represent the velocity vector field, and the contour (in black) represents a constant mass density of $10^{-14}$ kg m$^{-3}$.}
    \label{fig:Er_mag_Fr_evolution}
\end{figure*}

To show whether the radiation transport terms affect the formation, morphology, and evolution of the jet, in Figure~\ref{fig:Er_mag_Fr_evolution}, we display maps of the radiation energy density $E_{r}$ and the magnitude of the radiation flux $|{\bf F}_{r}|$ with the vector velocity field and a density contour (in red) of $10^{-14}$ kg m$^{-3}$ at three representative times, $t=100, 400, 800$ s. In addition, to schematize the dominant density energy in the evolution of the jet, at the bottom of Figure~\ref{fig:Er_mag_Fr_evolution}, we depict spatial profiles of the ratio $E_{r}/e_{\mathrm{int}}$ at the same times as $E_{r}$ and $|{\bf F}_{r}|$. For example, in the upper panels of Figure~\ref{fig:Er_mag_Fr_evolution} at the three times, the radiation energy density maps $E_{r}$ show a constant value in the region that covers the jet structure, which is delimited by the mass density contour in red. In the middle panels of Figure~\ref{fig:Er_mag_Fr_evolution}, we distinguish that the magnitude of the radiation flux $|{\bf F_{r}}|$ is low in the chromosphere, but increases at the corona. Despite this, the radiation flux remains constant in the region where the jet develops and does not shape the jet morphology either. In the bottom panels of Figure~\ref{fig:Er_mag_Fr_evolution}, the maps of the ratio $E_{r}/e_{int}$ show that $e_{\mathrm{int}}$ dominates the radiation energy density $E_{r}$ in the corona, but in the chromosphere and transition region the radiation energy density $E_{r}$ is more significant than $e_{\mathrm{int}}$. That said, the internal energy density is what gets the shape and morphology of the jet, since it is directly associated with the plasma variables, such as mass density and velocity. In all of the panels, the velocity vector field shows consistency in the jet's dynamics. Evidently, a strong negative flow at the solar corona ultimately diffuses the jet. Although the radiation transport variables do not get the shape and morphology of the jet, both affect their evolution, since they permeate the medium where the jet forms and propagates. In Appendix \ref{Appendix_RadMHD_blast_wave}, we show representative results of a basic test that show the behavior of the radiative version of the MHD blast wave considering the limits of an optically thick medium and an optically thin medium. This test illustrates how plasma fluid depends on the evolution of radiation transport equations in a more basic scenario than the study here.

The simulation results indicate that radiation transport contributes to the evolution of chromospheric jets, which is consistent with previous studies that have shown the key role of radiative cooling in shaping their properties \citep{Sterling&Mariska1990, 2000SoPh..196...79S, Gonzalez-Avilesetal2021}. Radiative losses influence the temperature, density, and velocity of the chromospheric jet by extracting thermal energy from the plasma, modifying pressure gradients, and the overall energy budget. This process, in turn, governs the height, lifetime, and morphology of the jets.

Furthermore, the results presented in this work exhibit notable similarities with those reported by \cite{Sterling&Mariska1990, Gonzalez-Avilesetal2021}, while also introducing a novel approach to modeling chromospheric jets. In \cite{Sterling&Mariska1990}, one-dimensional simulations assuming a pure hydrogen plasma with optically thin radiative losses and NLTE hydrogen ionization demonstrated that a substantial fraction of the injected energy is redistributed to ionization rather than heating. This redistribution leads to spicules that are shorter and weaker in their model. However, it is important to note that those simulations were performed in a hydrogen-only plasma, whereas in our study we adopt a more realistic composition of hydrogen and helium, corresponding to a mean molecular weight of $\mu=0.6$. The inclusion of helium modifies the thermodynamic response of the plasma by changing the specific heat, sound speed, and energy partition, which, in turn, can affect the jet’s evolution and energy transfer. Therefore, our comparison with hydrogen-only studies should be interpreted qualitatively: while the specific thermodynamic response may differ, the general conclusion that radiative processes redistribute energy away from heating and into ionization remains valid. In this context, the present work extends earlier findings by explicitly solving the two-moment radiation transport equations, thereby capturing the optically thin radiative cooling in a self-consistent way for a more realistic solar composition.

\subsection{Effects of the source terms on the background
solar atmosphere model}

\begin{figure*}
    \centering
    \includegraphics[width=8.0cm, height=6.0cm]{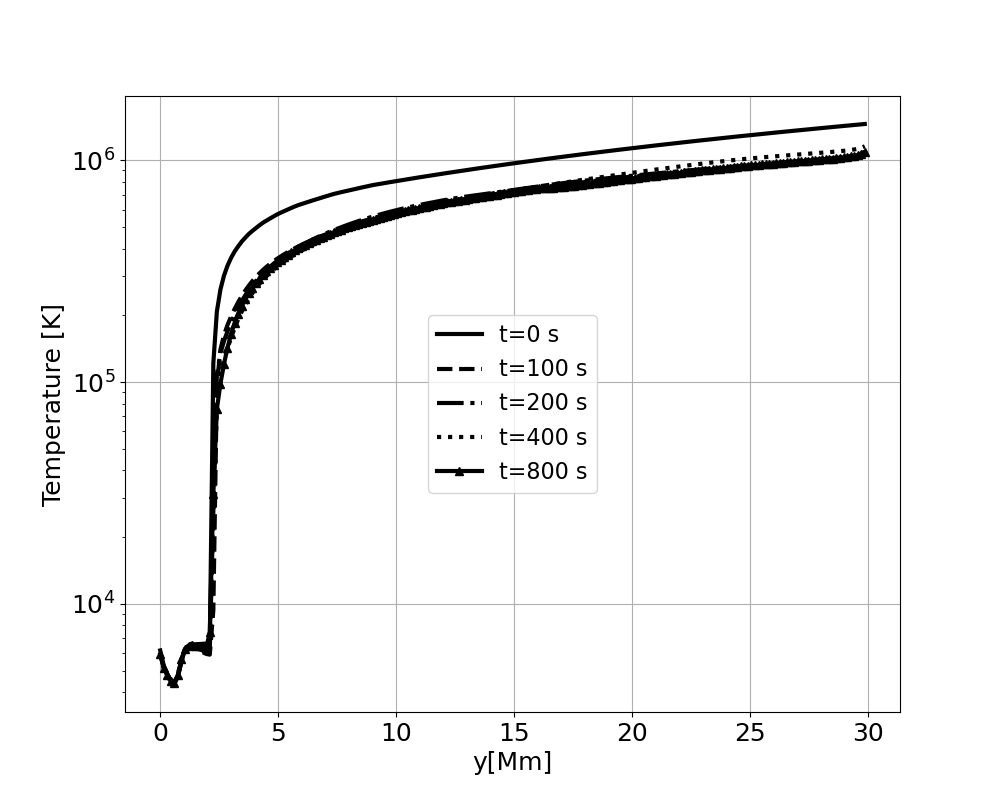}
    \includegraphics[width=8.0cm, height=6.0cm]{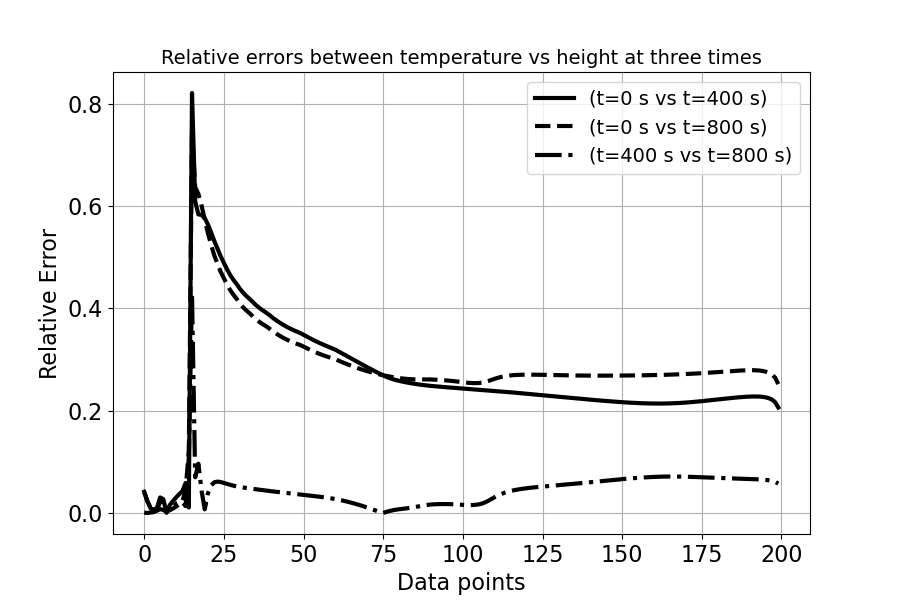}
    \caption{(Left) 1D cuts corresponding to five times ($t=0,100,200,400$ and 800 s) of the solar background model's plasma temperature in Kelvin as a function of height $y$ in Mm. (Right) The relative error between temperature values for three times $t=0,400$, and 800 s.}
    \label{fig:Temp_vs_height_times}
\end{figure*}

To qualitatively and quantitatively measure whether the temperature of the background atmosphere varies without the vertical velocity perturbation, but considering the non-ideal effects such as thermal conduction, coronal heating, and radiation, in Figure~\ref{fig:Temp_vs_height_times}, we display 1D cuts of the temperature as a function of the height $y$ at five different times ($t=0, 100, 200, 400, 800$ s). We also calculate the relative error from point to point between the temperature values at the initial time ($t=0$ s) compared to the temperature values at $t=400$ s and $t=800$ s and the relative error between the temperatures values at $t=400$ s and $t=800$. For example, on the left of Figure~\ref{fig:Temp_vs_height_times}, we depict the 1D temperature profiles on a vertical line traced from $y=0$ Mm to $y=30$ Mm, which is located in the middle of the numerical 2D domain, that is, at $x=15$ Mm. We can see that the temperature decreases slightly from time $t=100$ to the final time $t=800$ s. Despite the decrease in the background temperature profile, the variations are not substantial, and the solar corona temperature remains globally stable against the effect of the source terms. Furthermore, to quantitatively show the errors between the 1D temperature cuts, on the right of Figure \ref{fig:Temp_vs_height_times}, we show the results of the estimation of the relative error calculated according to the following expression, 

\begin{equation}
   \textrm{Relative error} = \left|\frac{T_{i}-T_{0}}{T_{0}}\right|, 
\end{equation}
where $T_{0}$ is the temperature profile at $t=0$ s, and $T_{i}$ with $i=1,2$ are the temperature profiles at $t=400$ s and $t=800$\,s, respectively. Therefore, in that plot, we observe a high relative error between $T_{0}$ and $T_{1}$ for some data points. This high value is also reached by the relative error between $T_{0}$ and $T_{2}$, and $T_{1}$ and $T_{2}$. Although this is a high value of the relative error, in the three curves, the relative error decreases and tends to a low value. The lowest relative error is between $T_{1}$ and $T_{2}$, while the highest relative error is achieved between $T_{0}$ and $T_{1}$, but it is of the same order as $T_{0}$ and $\mathbf{T_{2}}$. The high value of the relative error in the three comparisons is due to the values at the transition region, since a sharp gradient could make a point-to-point comparison vary significantly.      

\section{Conclusions and final comments}
\label{sec:conclusions_and_final_comments}

The main objective of this paper was to study the formation and evolution of an impulsively driven chromospheric jet that resembles some spicule features using a 2D non-ideal radiation MHD numerical simulation. Specifically, we model the solar atmosphere and the jet as a plasma that follows the MHD equations coupled to a system of frequency- and angle-averaged radiation transport equations that include the evolution of the radiation energy flux and the radiation flux components. To trigger the formation of the jet, we perturb the solar atmosphere with a localized Gaussian pulse in the vertical velocity, set at the solar chromosphere ($y=1.75$ Mm). In addition, we initially considered a solar atmosphere in non-local thermodynamic equilibrium (NLTE) by setting characteristic low values of the Planck-averaged absorption opacity ($\kappa=1\times10^{-6}$ cm$^{2}$ g$^{-1}$) and the Rosseland-averaged scattering opacity ($\sigma=0.34$ cm$^{2}$ g$^{-1}$) corresponding to an optically thin medium such as the solar corona.   

Based on the results of the numerical simulation, we observe classical features on the mechanism of jet formation reported in various numerical MHD simulations of impulsively driven jets \cite{Muraswki&Zaqarashvili2010, Muraswkietal2011, Gonzalez-Avilesetal2021, Srivastavaetal2023}, where the mechanism can be explained in terms of the following physics. The localized pulse rapidly transforms into shocks that propagate upward and generate the collimated plasma structure with some features of a solar spicule. In addition, the pulse also excites a non-linear wake in the chromosphere, which leads to quasi-periodic secondary shocks, which in turn lift the chromospheric plasma upward and create quasi-periodic jets in the solar corona. In summary, the chromospheric jet modeled in this study exhibits some characteristics that resemble observed macrospicules, while also presenting differences. In terms of height, the simulated jet reaches a maximum vertical extent of approximately 8 Mm above the transition region, placing it at the lower limit of the observed macrospicule height range, which spans from 7 to 70 Mm above the solar limb \citep{2000A&A...360..351W}. The width of the simulated jet, approximately 1 Mm, is also consistent with the typical macrospicule widths, which are generally around 3 Mm. However, the maximum vertical velocity reached in the simulation, about 40 km s$^{-1}$, falls below the lower bound of the observed velocity range for macrospicules, which lies between 70 and 140 km s$^{-1}$ \citep{Loboda_2019}. This suggests that the simulated jet corresponds to a relatively low-energy event. Similarly, the maximum temperature attained in the simulation, approximately $8 \times 10^4$ K, is lower than the typical macrospicule temperature range of $(1$–$2) \times 10^5$ K \citep{1991ApJ...376L..25H}. Despite these differences, the simulation successfully reproduces several dynamical features commonly associated with macrospicules, including shock formation, double-thread structures, and jet dissipation. The inclusion of radiation transport equations that evolve both the radiation energy density and flux demonstrates that radiation acts as an efficient cooling mechanism in the optically thin corona. This cooling facilitates downward flows and contributes to the eventual dissipation of the jet. The radiative behavior observed in this study is consistent with the results of earlier models employing empirical radiative loss functions \citep[e.g.,][]{Gonzalez-Avilesetal2021}, reinforcing the conclusion that radiative losses, whether empirically prescribed or self-consistently computed by radiation transport, play a central role in limiting the height and energy content of macrospicules.

Concerning the effects of the radiation transport term on the chromospheric jet, we identify that either the radiation energy density or radiation flux determines the morphology of the jet, but both variables permeate the solar atmospheric environment where the jet propagates. In particular, the radiation energy density distributes more homogeneously than radiation energy flux over the solar atmosphere domain without significantly affecting the background solar model. Interestingly, the radiation energy density and radiation flux at the solar corona lead to a negative flow that diffuses the jet and makes it tiny. This result is consistent with the result obtained in \cite{Gonzalez-Avilesetal2021}, where the authors explored the effects of radiative cooling on the formation and evolution of solar jets with some macrospicule features. In that paper, the authors found that radiative cooling produced a negative flow that dissipated the jet, making it short and cold. Such a study considers radiative cooling represented by an optically thin radiative loss function computed according to the CHIANTI code, which is an atomic database and software package used to model emission lines and continua from hot and optically thin astrophysical plasmas \cite{1997A&AS..125..149D}. The current study did not consider the optically thin radiative loss function as a source term in the total energy density equation. Instead, we considered the coupling system of the MHD equations with the radiation transport equations that had never been attempted before. We found that in an optically thin region such as the solar corona, where the scattering and absorption opacities are low, the radiation terms act as an optically thin radiative loss mechanism. This result is interesting since it tells us that the more appropriate treatment of the coupling between matter and radiation in the context of spicule simulations could be the solution of the MHD equations with radiative transfer, which includes the optically thick radiation and the numerical recipes for the radiative transfer similar to what has been done in, for example, \cite{Carlsson&Leenaarts2012,Martinez-Sykoraetal2017a, Martinez-Sykoraetal2017b}. Finally, in this paper, it is the first time so far that the evolution of the radiative energy density and radiative flux in the formation and evolution of an impulsively driven chromospheric jet has been shown and deserves attention to explore potential applications in future studies in optically thick regions such as solar convection phenomena.

\section*{Acknowledgments}

The author acknowledges the ``Secretar\'ia de Ciencia, Humanidades, Tecnolog\'ia e Innovaci\'on (SECIHTI)'' 319216 project ``Modalidad: Paradigmas y Controversias de la Ciencia 2022'', and the "Programa de Apoyo a Proyectos de Investigaci\'on e Innovaci\'on Tecnol\'ogica (PAPIIT)" project IA100725 for supporting the investigation of this work. The author thanks the developers of the PLUTO software, which provides a general and sophisticated interface for the numerical solution of mixed hyperbolic/parabolic systems of partial differential equations (conservation laws) targeting high Mach number flows in astrophysical fluid dynamics \citep{Mignoneetal2007}. In addition, the author expresses sincere gratitude to the referee for the careful and thorough review of the manuscript. The suggestions, corrections, and comments provided are highly appreciated and have contributed significantly to improving the overall quality and clarity of the article. Finally, the author acknowledges the use of the NumPy package \citep{harris2020array} for numerical computations and Matplotlib \citep{hunter2007matplotlib} for data visualization.

\appendix

\section{Dynamics of the plasma and the radiation field in a basic MHD test}
\label{Appendix_RadMHD_blast_wave}

To illustrate how the plasma behaves while coupling with the radiation field, in this appendix \ref{Appendix_RadMHD_blast_wave}, we show representative results of the radiation version of the MHD blast wave test for optically thick and optically thin cases. In the ideal MHD equations, the blast wave measures the ability of the numerical algorithms implemented in the PLUTO code to handle strong waves propagating in highly magnetized environments. It is also helpful to test the code to control nonphysical densities or pressures by adequately controlling the divergence-free condition. The 2D radiative version of the MHD blast wave consists of the same setup for the plasma variables as for the ideal MHD test. However, it includes initial conditions for the radiation energy density and the radiation flux (\ref{mass_cons_eq}-\ref{flux_rad_eq}), but without dissipative terms in a 2D Cartesian domain. 

\begin{figure*}
    \centering
    \includegraphics[width=6.0cm, height=4.0cm]{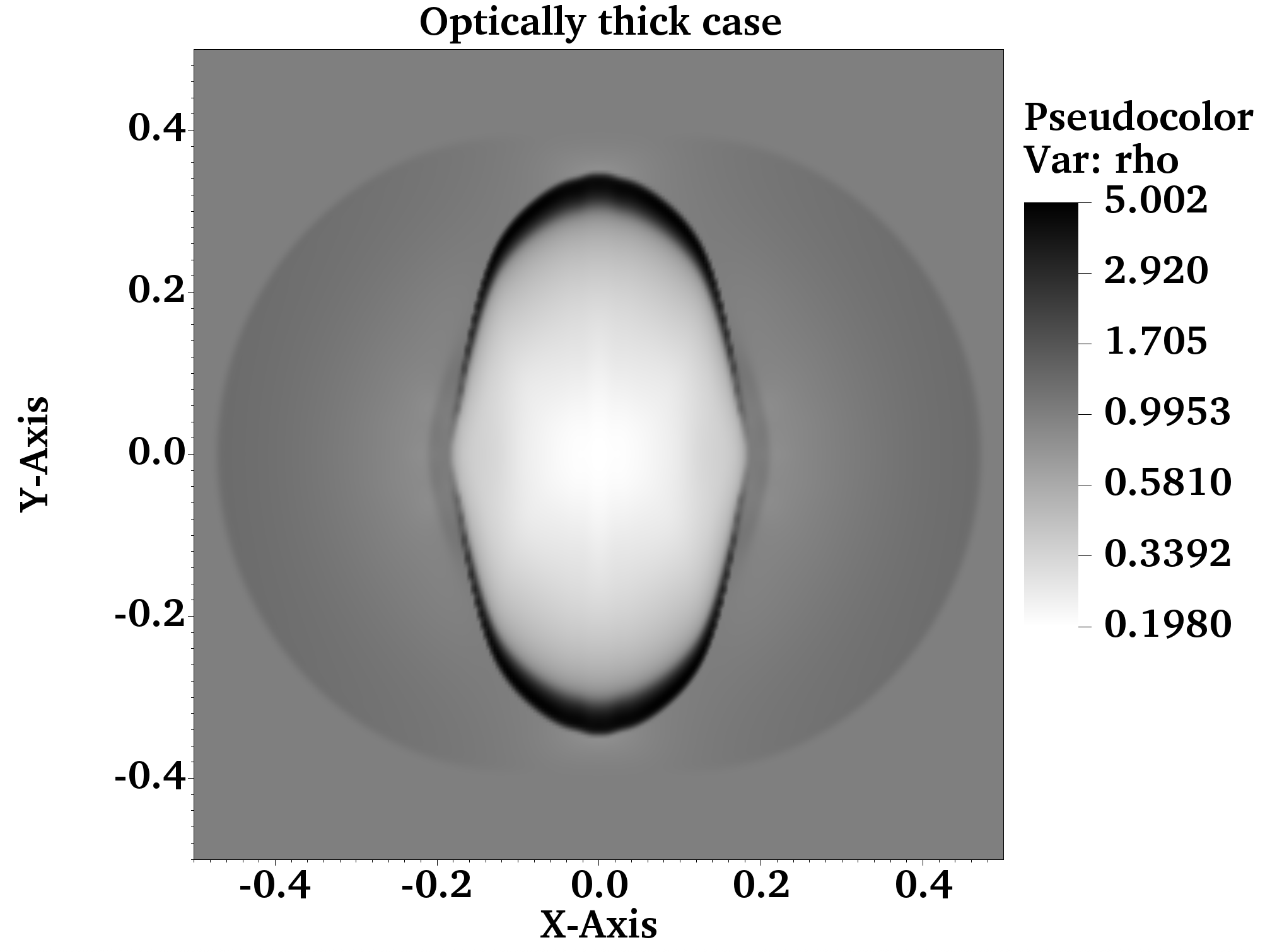}
    \includegraphics[width=6.0cm, height=4.0cm]{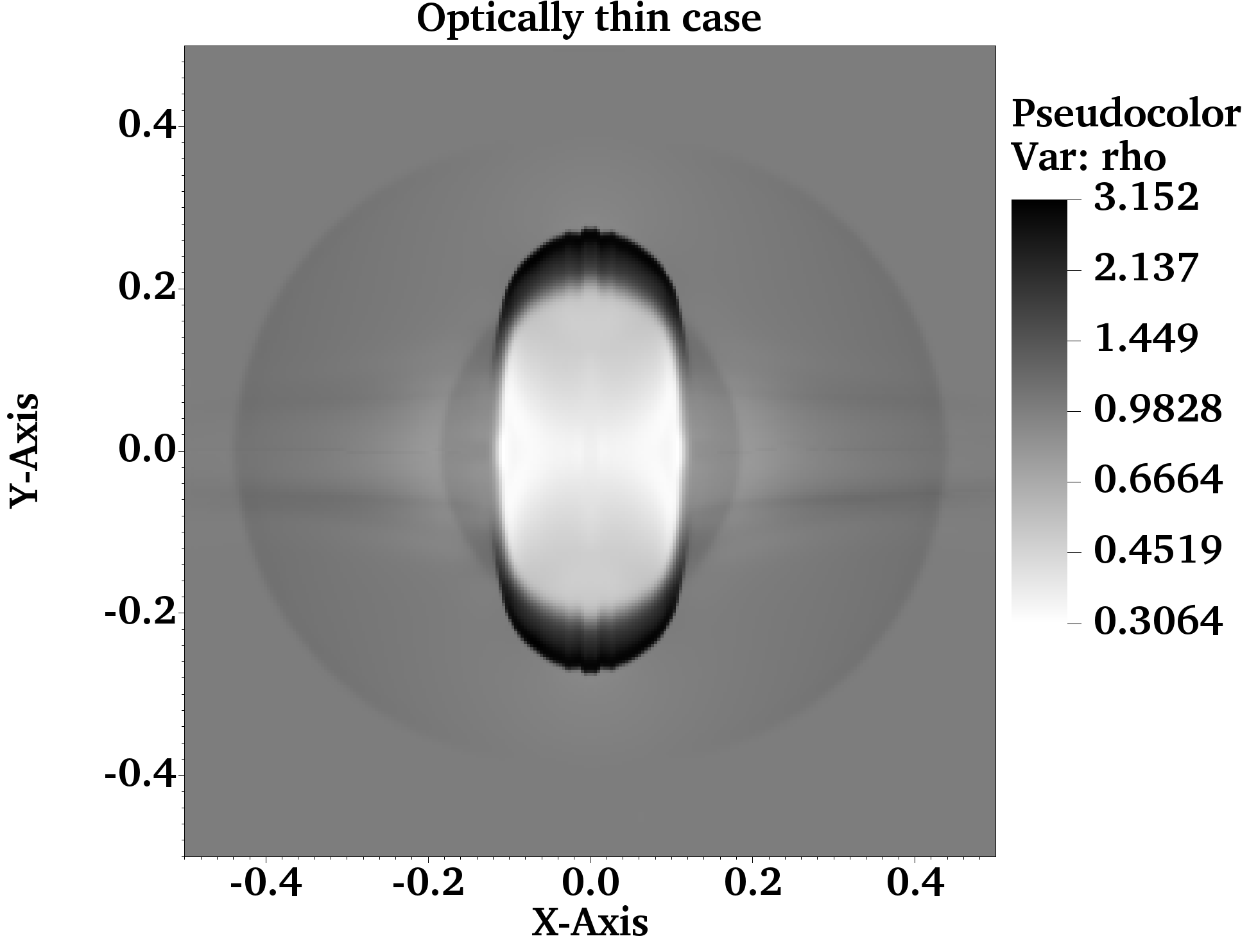} \\
    \includegraphics[width=6.0cm, height=4.0cm]{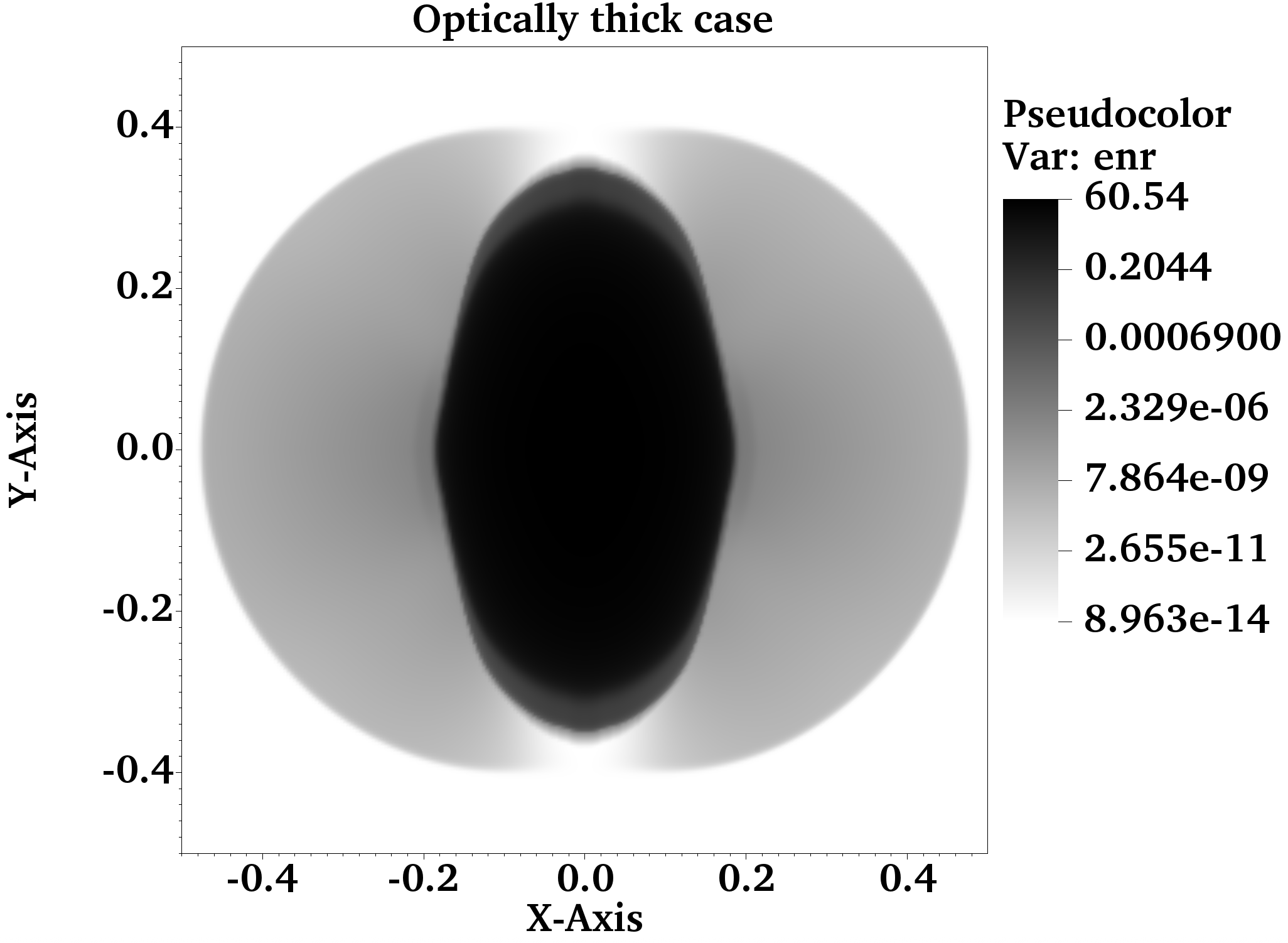}
    \includegraphics[width=6.0cm, height=4.0cm]{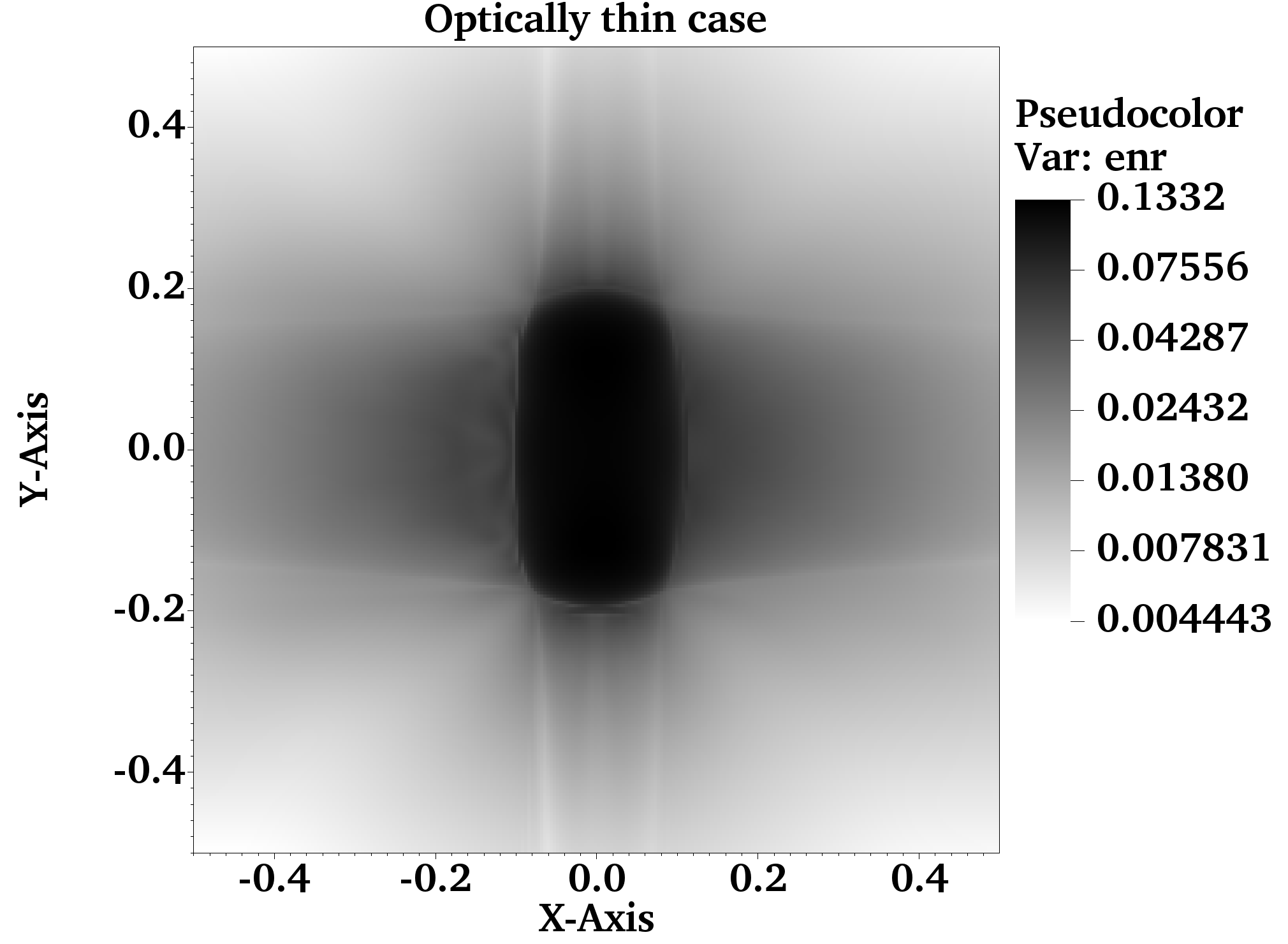} \\
    \includegraphics[width=6.0cm, height=4.0cm]{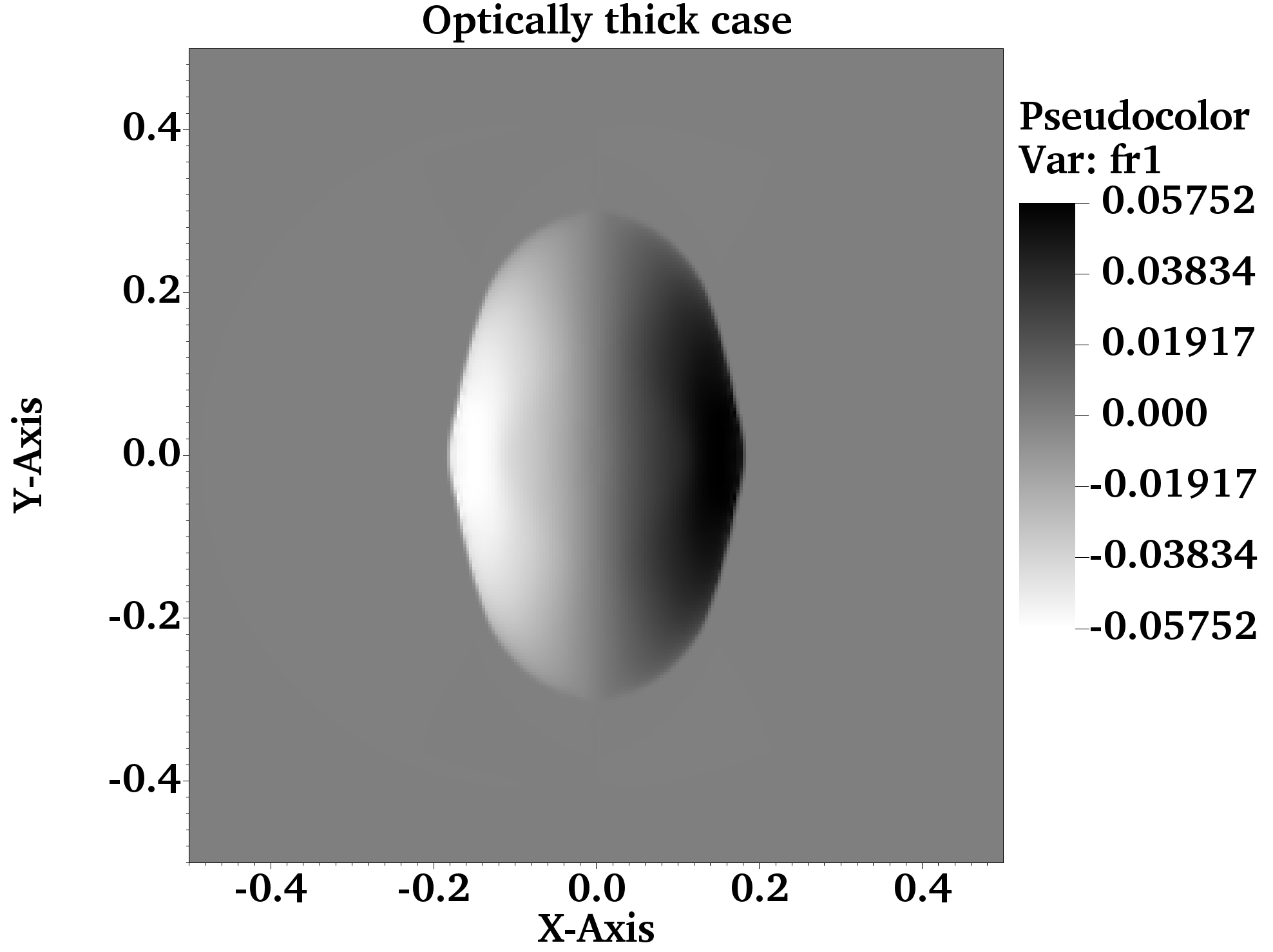}
    \includegraphics[width=6.0cm, height=4.0cm]{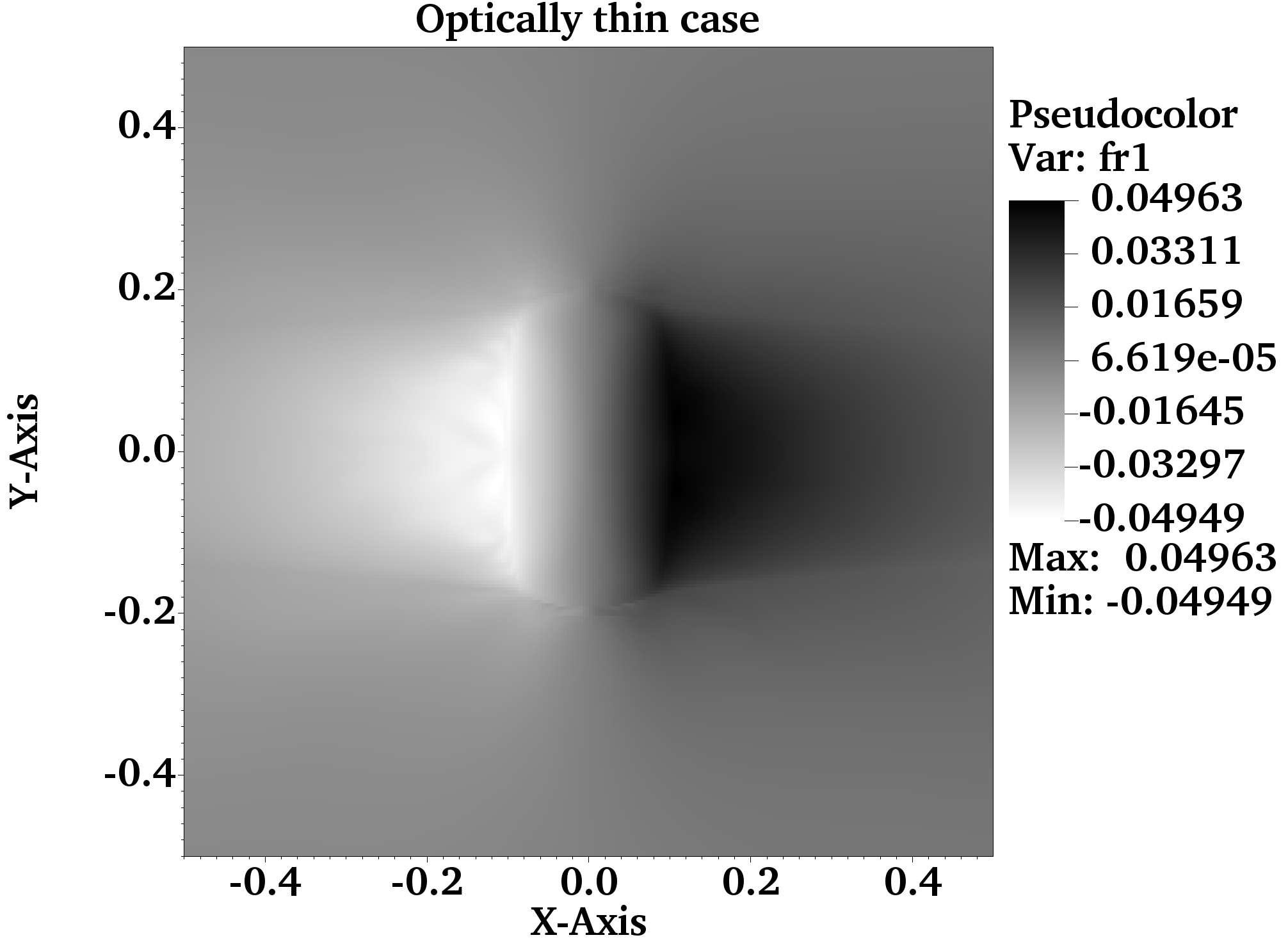} 
    \caption{Spatial profiles of the mass density (top), radiation energy density (middle), and radiation energy flux (bottom) correspond to the 2D radiative MHD blast wave test for the optically thick case (left) and optically thin case (right). All the plots are at $t=0.01$ in code units.}
    \label{fig:Rad_MHD_test}
\end{figure*}

In Figure \ref{fig:Rad_MHD_test}, we show maps of the mass density (top), radiation energy density (middle), and the $x-$component of the radiation flux (bottom) for the optically thick (left) and optically thin (right) scenarios. For the optically thick case, that is, for a high absorption opacity ($\kappa=10^{4}$), we see the typical behavior of the mass density, similar to the blast wave in the ideal MHD approximation. However, interestingly, the radiation energy density shows a morphology similar to that of the mass density, and it also contributes to the acceleration of the MHD shock wave. Furthermore, the $x-$ component of the radiation flux also couples with the evolution of the plasma mass density. This behavior implies that the plasma is predominantly coupled with the radiation field in the optically thick scenario.
On the other hand, in the right panels of Figure \ref{fig:Rad_MHD_test}, we depict the maps for the optically thin scenario, that is, for a low absorption opacity ($\kappa=1$). In these panels, it is clear that the morphology of the mass density is different from that in the optically thick case. Precisely, the plasma density moves slowly at the front shock, and the radiation energy density is low, which does not significantly affect the plasma motion. In addition, the map of the $x-$ component of the radiation flux schematizes the propagation opposite to the intense shock front.          

\end{multicols}

\medline
\begin{multicols}{2}
%
\nocite{*}
\bibliographystyle{rmf-style}
\bibliography{Radiation_jets_RMF_2025}
\end{multicols}
\end{document}